\documentclass[12pt,leqno]{article}
\title{Thermodynamics of self-gravitating systems}
\author{Pierre-Henri Chavanis$^{1,2}$, Carole Rosier$^{3}$ and Cl\'ement Sire$^{1}$}
\date{}

\usepackage{amsthm,amsmath,amssymb,a4wide,psfig,epsf}
\def\mb#1{\setbox0=\hbox{$#1$}\kern-.025em\copy0\kern-\wd0
\kern-0.05em\copy0\kern-\wd0\kern-.025em\raise.0233em\box0}

\begin{document}
\maketitle
\vspace*{-1cm}
\begin{center}
$^{1}$ Laboratoire de Physique Quantique - UMR CNRS 5626,
Universit\'e Paul Sabatier,\\
118, route de Narbonne 31062 Toulouse, France.\\

$^{2}$ Institute for Theoretical Physics,
University of California, Santa Barbara, California.\\

$^{3}$ UMR CNRS 5585, Analyse Num\'erique,
Universit\'e Lyon 1, b\^at. 101,\\
69622 Villeurbanne Cedex, France. \\
\vspace{0.5cm}
\end{center}

\begin{abstract}

We study the thermodynamics and the collapse of a self-gravitating gas
of Brownian particles. We consider a high friction limit in order to
simplify the problem. This results in the Smoluchowski-Poisson
system. Below a critical energy or below a critical temperature, there
is no equilibrium state and the system develops a self-similar
collapse leading to a finite time singularity.  In the microcanonical
ensemble, this corresponds to a ``gravothermal catastrophe'' and in the
canonical ensemble to an ``isothermal collapse''. Self-similar
solutions are investigated analytically and numerically.

\end{abstract}

\section{Introduction}
\label{sec_introduction}

The thermodynamics of self-gravitating systems displays intriguing
features due to the existence of negative specific heats,
inequivalence of statistical ensembles and phase transitions
associated with gravitational collapse \cite{pad}. Thermodynamical
equilibrium of a self-gravitating system enclosed within a box exists
only above a critical energy $E_{c}=-0.335GM^{2}/R$ or above a
critical temperature $T_{c}=GMm/ 2.52 kR$ and is at most a {metastable}
state, i.e. a {\it local} maximum of a relevant thermodynamical
potential (the entropy in the microcanonical ensemble and the free
energy in the canonical ensemble) \cite{antonov,lbw}. For $T<T_{c}$ or
$E<E_{c}$, the system is expected to collapse. This is called
``gravothermal catastrophe'' or ``Antonov instability'' in the
microcanonical ensemble (MCE) and ``isothermal collapse'' in the
canonical ensemble (CE). Dynamical models appropriate to star
formation \cite{penston} or globular clusters
\cite{larson,cohn,lbe,lance} show that the collapse is self-similar
and leads to a finite time singularity (i.e., the central density
becomes infinite in a finite time). The value of the scaling exponent
in the density profile $\rho\sim r^{-\alpha}$ depends whether the
system evolves at fixed temperature (in which case $\alpha=2$ results
from dimensional analysis) or if its temperature is free to diverge
(in which case the value of the exponent is non trivial and often
close to $2.2$). It is found in general that the shrinking of the core
is so rapid that the core mass goes to zero at the collapse time
although the central density is infinite.

In this paper, we introduce a simple model of gravitational
dynamics which exhibits similar features and which can be studied
in great detail.  Specifically, we consider a gas of
self-gravitating Brownian particles enclosed within a spherical
box. For simplicity, we take a high friction limit and reduce the
problem to the study of the Smoluchowski-Poisson system. In the
simplest formulation, the temperature is constant (canonical
description). We also consider the case of an isolated medium with an
infinitely large thermal conductivity so that its temperature is
uniform in space but varies with time in order to conserve energy
(microcanonical description). The interest of these models is
their relative simplicity which allows for a complete theoretical
analysis, while keeping all the richness of the thermodynamical
problem: inequivalence of statistical ensembles, phase
transitions, gravitational collapse, finite time singularity,
persistence of metastable states, basin of attraction... These
models are consistent with the first and second principles of
thermodynamics and give a dynamical picture of what happens when
no equilibrium state exists. However, in view of their
considerable simplification, it is not clear whether these models
can have astrophysical applications although connections with the
dynamics of dust particles in the solar nebula and the process of
``violent relaxation'' in collisionless stellar systems are
mentioned.

The paper is organized as follows. In Sec. \ref{sec_smoluchowski}, we
introduce the Smoluchowski-Poisson (SP) system for a gas of
self-gravitating Brownian particles and list its main properties. In
particular, we make contact with thermodynamics and show that the SP
system satisfies a form of $H$-theorem.  In Sec. \ref{sec_statmech},
we discuss the existence of stationary solutions of the SP system and
the relation with maximum entropy states. In Sec. \ref{sec_stab}, we
perform a linear stability analysis of the SP system. We show that a
stationary solution is linearly stable if and only if it is a local
entropy maximum and that the eigenvalue problem for linear stability
is connected to the eigenvalue problem for the second order variations
of entropy studied in Refs. \cite{pad2,chavcano}. In
Sec. \ref{sec_selfsimilar}, we consider the case of gravitational
collapse and exhibit self-similar solutions of the SP system. Since
the particles are confined within a box, there is a small deviation to
the purely self-similar regime and we describe this correction in
detail.

In Sec. \ref{sec_simulations}, we perform various numerical simulations
of the SP system for different initial conditions. We check the
results of thermodynamics, namely the existence of equilibrium states
for $E>E_{c}$ and $T>T_{c}$ and the gravitational collapse otherwise.
 We find that the collapse proceeds
self-similarly with explosion, in a finite time $t_{coll}$, of the
central density while the core radius shrinks to zero. In MCE this is
accompanied by a divergence of temperature and entropy. In the
limit $t\rightarrow t_{coll}$, we find the scaling laws
$\rho_{0}r_{0}^{\alpha}\sim 1$ and $\rho/\rho_{0}\sim
(r/r_{0})^{-\alpha}$. The scaling exponent is $\alpha=2$ in CE and
$\alpha\simeq 2.21$ in MCE. In CE, the invariant profile
$\rho/\rho_{0}=f(r/r_{0})$ can be determined analytically. The
collapse time diverges like $t_{coll}\sim (E_{c}-E)^{-1/2}$ and
$t_{coll}\sim (T_{c}-T)^{-1/2}$ as we approach the critical energy
$E_{c}$ and critical temperature $T_{c}$. We also study the linear
development of the instability (for unstable isothermal spheres) and
show that the density perturbation $\delta
\rho/\rho$ presents several oscillations depending on the value of the
density contrast. In particular, at the points of marginal stability
in the series of equilibria, the perturbation $\delta \rho/\rho$ has a
``core-halo'' structure in the microcanonical ensemble but not in the
canonical ensemble in agreement with theory
\cite{pad2,chavcano}.

\section{Self-gravitating Brownian particles}
\label{sec_brown}

\subsection{The Smoluchowski-Poisson system}
\label{sec_smoluchowski}

We consider a system of small particles with mass $m$ immersed in a
fluid. We assume that the fluid imposes to the particles a friction
force $-\xi {\bf v}$ and a stochastic force ${\bf R}(t)$. This random
force may mimic ordinary Brownian motion (i.e. the collisions of the
fluid particles onto the solid particles) or fluid turbulence. We
assume in addition that the particles interact gravitationally with
each other. Therefore, the stochastic Langevin equation describing the
motion of a particle reads
\begin{equation}
{d{\bf v}\over dt}=-\xi {\bf v}+{\bf F}({\bf r},t)+{\bf R}(t),
\label{X0}
\end{equation}
where ${\bf F}=-\nabla\Phi$ is the gravitational force acting on the
particle. For simplicity, we shall assume that the stochastic force is
delta-correlated in time and set
\begin{equation}
\langle {\bf R}(t){\bf R}(t')\rangle=6 D \ \delta (t-t'),
\label{X1}
\end{equation}
where $D$ measures the noise strength of the Langevin force. In order to
recover the Maxwell-Boltzmann distribution
\begin{equation}
{f}={1\over (2\pi T)^{3/2}}\rho e^{-{v^{2}\over 2 T}} \qquad {\rm with}
\qquad \rho=A e^{-\beta \Phi},
\label{fMB}
\end{equation}
at equilibrium, the diffusion coefficient and the friction term
must be related according to the Einstein relation $D={\xi T}$.
Applying standard methods \cite{risken}, we can immediately write
down the Fokker-Planck equation associated with this stochastic
process:
\begin{equation}
{\partial {f}\over\partial t}+{\bf v}{\partial {f}\over\partial {\bf
r}}+{\bf F}{\partial {f}\over\partial {\bf v}}={\partial\over\partial
{\bf v}}\biggl \lbrace D\biggl ({\partial {f}\over\partial {\bf
v}}+\beta {f}{\bf v}\biggr )\biggr\rbrace.
\label{X4}
\end{equation}
This is the familiar Kramers equation but, when self-gravity is taken
into account, it must be coupled to the Poisson equation
\begin{equation}
\Delta \Phi =4\pi G \rho,
\label{Poisson}
\end{equation}
where $G$ is the gravitational constant. This makes its study much
more complicated than usual. The Kramers-Poisson (KP) system was first
introduced in astrophysics by Chandrasekhar \cite{sto} in his
stochastic theory of stellar dynamics (for, e.g., globular
clusters). In that context, the diffusion and the friction arise
self-consistently as the result of the fluctuations of the
gravitational field. An equation of the form (\ref{X4}) was also
proposed as an {\it effective} dynamics of collisionless stellar
systems (on a coarse-grained scale) during the period of violent
relaxation \cite{lb,quasi}.

In order to simplify the problem, in a first approach, we consider a high
friction limit $\xi\rightarrow +\infty$.  Then, it is possible to
neglect the inertial term in the Langevin equation (\ref{X0}). The
Fokker-Planck equation describing this high friction limit is the
Smoluchowski equation
\begin{equation}
{\partial\rho\over\partial t}=\nabla\biggl \lbrace {1\over
\xi}(T\nabla \rho+\rho\nabla\Phi)\biggr\rbrace,
\label{X5}
\end{equation}
with a diffusion coefficient $D'=T/\xi$ and a drift term proportional
to the gravitational force.  The ordinary Smoluchowski equation
describes the sedimentation of colloidal suspensions in an external
gravitational field. Since it is a prototype of kinetic equations, it
is clearly of great interest to consider the extension of this model
to the case where the potential is not fixed but related to the
density of the particles via a Poisson equation, like in the
gravitational case.

The Smoluchowski equation can be interpreted equivalently as a
continuity equation for the density $\rho$ with a velocity field
\begin{equation}
{\bf u}=-{1\over\xi}\biggl ({T\over\rho}\nabla\rho+\nabla\Phi\biggr ),
\label{X6}
\end{equation}
where $-T\nabla\rho$ is the pressure force and $-\rho\nabla\Phi$ the
gravitational force. At equilibrium, the two terms balance each other
and the Boltzmann distribution (\ref{fMB}) establishes
itself. Physically, the high friction limit supposes that there are
two time scales in the problem. On a short time scale of the order of
the friction time $\xi^{-1}\ll t_{dyn}$, the system thermalizes and
the distribution function becomes Maxwellian with temperature $T$
(this is obvious if we take the limit $D=\xi T\rightarrow +\infty$ in
the r.h.s. of Eq. (\ref{X4})). Then, on a longer time scale of the
order of the dynamical time $t_{dyn}$, the particle distribution
$\rho({\bf r},t)$ tends to evolve towards a state of mechanical
equilibrium described by the Boltzmann distribution (\ref{fMB}). Note
that the opposite assumptions are made for globular clusters
\cite{larson,cohn,lbe}: the system is assumed to be in
mechanical equilibrium and the evolution is due to thermal transfers
between the core and the halo. Our model of self-gravitating Brownian
particles could find applications for the dynamics of dust particles
in the solar nebula and the formation of planetesimals by
gravitational instability (see, e.g., Ref. \cite{caa}). In that
context, the dust particles experience a friction with the gas modeled
by Stokes or Epstein's laws and the high friction limit may be
relevant. On the other hand, the diffusion of the particles could
result from a stochastic component of the force or from fluid
turbulence. This would be just a
first approach because the physics of planetesimal formation is
more involved than our simple model.

Since the system described previously is in contact with a heat bath,
the proper statistical treatment is the {\it canonical ensemble} in
which the temperature $T$ is fixed. In order to test dynamically the
inequivalence of statistical ensembles for self-gravitating systems,
we would like to introduce a simple model corresponding to the {\it
microcanonical ensemble}, i.e. with strict conservation of energy
$E$. In fact, when a Brownian particle moves with its terminal
velocity in a gravitational field, the work of the force ought to be
converted into heat. If the medium acts as a thermostat with an
infinite volume and with rapid dissipation of heat, we can disregard
the variation of temperature and we get the isothermal model discussed
previously. However, if we are to keep track of local heating, the
temperature will depend on space and time and we need to set up a
model in which energy is conserved. Such a generalization of Brownian
theory has recently been developed by Streater
\cite{streater} in the case of an external gravitational potential. This
{\it statistical dynamics} approach \cite{sd} leads to coupled
nonlinear equations for the density $\rho({\bf r},t)$ and the
temperature $T({\bf r},t)$ which are consistent with the first and
second principles of thermodynamics.  Such equations can be {derived}
from a microscopic model involving Brownian particles and heat
particles modeled as quantum oscillators. A generalization of these
equations for self-gravitating Brownian particles has been proposed by
Biler {\it et al.} \cite{biler}. It consists of the
Smoluchowski-Poisson system (\ref{X5}) (\ref{Poisson}) coupled to a
diffusion equation for the temperature
\begin{equation}
{3\over 2}{\partial\over\partial t}(\rho T)=\nabla(\lambda\nabla T)-
\nabla(T{\bf J})-{\bf J}\nabla\Phi,
\label{streater}
\end{equation}
where ${\bf J}$ is the diffusion current in Eq. (\ref{X5}).
However, this model still remains complicated for a first
approach. Since our main purpose is to illustrate in the simplest
way the basic features of the thermodynamics of self-gravitating
systems (inequivalence of ensembles, gravothermal catastrophe,
isothermal collapse, phase transitions, basin of attraction...),
we shall consider an additional approximation and let the thermal
conductivity $\lambda$ in Eq. (\ref{streater}) go to $+\infty$. In
that case, the temperature is uniform but still evolving with time
according to the law of energy conservation (first principle):
\begin{equation}
E={3\over 2}M T(t)+{1\over 2}\int \rho\Phi \,d^{3}{\bf r}.
\label{EE2}
\end{equation}
The first term in the r.h.s is the kinetic energy $K=\int f{v^{2}\over
2}d^{3}{\bf r}d^{3}{\bf v}$ for a Maxwellian distribution function
with temperature $T$ (local thermodynamical equilibrium) and the
second term is the gravitational energy of interaction. Equations
(\ref{X5}) (\ref{Poisson}) (\ref{EE2}) lead to a simple microcanonical
model for self-gravitating systems with a lot of attractive
properties. The Cauchy problem for this system of equations was
studied by Rosier \cite{rosier}. These equations were first proposed
by Chavanis {\it et al.} \cite{csr} as a simplified model of ``violent
relaxation'' by which a stellar system initially far from mechanical
equilibrium tries to reach an isothermal state on a few dynamical
times \cite{lb,dubrovnik}. In that context, the engine of the evolution is the
competition between pressure and gravity, like in Eq. (\ref{X5}). This
particular equation corresponds to an overdamped evolution but more
general equations taking into account inertial terms are also proposed
in Ref. \cite{csr}.

It is easy to show that the SP system admits a form of $H$-theorem for
an appropriate thermodynamical potential (second principle). The
microcanonical ensemble is characterized by the specification of mass
$M$ and energy $E$. The thermodynamical potential is the entropy
\begin{equation}
S={3\over 2}M+{3\over 2}M\ln (2\pi T)-\int {\rho}\ln {\rho} \,d^{3}{\bf r},
\label{SS2}
\end{equation}
which is the form of the classical Boltzmann entropy $S=-\int f\ln
f d^{3}{\bf r}d^{3}{\bf v}$ for a Maxwellian distribution function
with temperature $T$. Then, it is easy to show, using Eqs.
(\ref{X5}) and (\ref{EE2}) that \cite{csr}:
\begin{equation}
\dot S=\int {1\over T \rho \xi}(T\nabla \rho+\rho\nabla\Phi)^{2}
\,d^{3}{\bf r}\ge 0.
\label{dotSeq}
\end{equation}
Therefore, the entropy plays the role of a Lyapunov function for our
microcanonical model.  The canonical ensemble is characterized by the
specification of mass $M$ and temperature $T$. It is straightforward
to show that the SP system (\ref{X5}) satisfies a relation similar to
Eq. (\ref{dotSeq}) for the free energy (more precisely the Massieu
function) $J=S-\beta E$. It can be noted that the Kramers equation
(\ref{X4}) and the Smoluchowski equation (\ref{X5}) can also be
derived from a variational formulation \cite{csr}, called the Maximum
Entropy Production Principle (M.E.P.P.). This makes a direct relation
between the dynamics and the thermodynamics. Since the SP system with
the constraint (\ref{EE2}) obeys the same conservation laws and
H-theorem as more realistic models such as Landau-Poisson system
\cite{lance} and coarse-grained Vlasov-Poisson system \cite{quasi}, it
should exhibit qualitatively similar properties even if the details of
the evolution are expected to differ in many respects.

To properly define our system of equations, we must specify the
boundary conditions. We shall assume that the system is non rotating
and restrict ourselves to spherically symmetric solutions. In
addition, we shall work in a spherical box of radius $R$ to avoid the
well-known infinite mass problem associated with isothermal
configurations.  In that case, the boundary conditions are:
\begin{equation}
{\partial\Phi\over\partial r}(0)=0,\qquad  \Phi(R)=-{GM\over R},
\qquad T{\partial \rho\over \partial r}+\rho {GM\over R^{2}}=0.
\label{boundary1}
\end{equation}
The first condition expresses the fact that the gravitational force at
the center of a spherically symmetric system is zero. The second
condition defines the gauge constant in the gravitational
potential. Finally, the last condition insures that the total mass is
conserved (we have used the Gauss theorem $\partial_{r}\Phi=GM/r^{2}$ to
simplify its expression).

For spherically symmetric systems, it is possible to reduce the SP
system to a single partial differential equation for the mass
profile $M(r,t)=4\pi\int_{0}^{r}\rho r^{'2}dr'$. Multiplying both
sides of Eq. (\ref{X5}) by $r^{2}$ and integrating from $0$ to $r$
we obtain after straightforward algebra
\begin{equation}
{\partial M\over\partial t}(r,t)={1\over\xi}\biggl\lbrace
T{\partial^{2}M\over\partial r^{2}}(r,t)-{2T\over r}{\partial
M\over\partial r}(r,t)+{GM(r,t)\over r^{2}}{\partial
M\over\partial r}(r,t)\biggr\rbrace. \label{massint1}
\end{equation}
The appropriate boundary
conditions are now $M(0,t)=0$ and $M(R,t)=M$. The potential energy can
be expressed in terms of $M(r,t)$ as \cite{bt}:
\begin{equation}
W=-\int_{0}^{R}{GM(r,t)\over r}{\partial M\over\partial r}(r,t)dr.
\label{massint4}
\end{equation}
It is possible to simplify Eq. (\ref{massint1}) a little more by
introducing the new coordinate $u=r^{3}$ so that
\begin{equation}
\xi{\partial M\over\partial t}(u,t)=9Tu^{4/3}{\partial^{2}M\over\partial
u^{2}}(u,t)+{3GM(u,t)}{\partial M\over\partial u}(u,t).
\label{massint1bis}
\end{equation}

Finally, we note that the Krammers-Poisson (KP) system satisfies a form  of Virial theorem:
\begin{equation}
{1\over 2}{d^{2}I\over dt^{2}}+{1\over 2}\xi {dI\over dt}=2K+W-3p_{b}V,
\label{virial5}
\end{equation}
where $I=\int \rho r^{2}d^{3}{\bf r}$ is the moment of inertia (we
have properly taken into account the pressure on the box). The
difference with the usual Virial theorem is the occurrence of a
damping term ${1\over 2}\xi\dot I$ due to friction.   In the high
friction limit, we get
\begin{equation}
{1\over 2} {dI\over dt}={1\over\xi}(2K+W-3p_{b}V).
\label{virial6}
\end{equation}
This expression can also be directly obtained from the SP system.

\subsection{Stationary solutions and maximum entropy states}
\label{sec_statmech}

The stationary solutions of the SP system are given by the
Boltzmann distribution (\ref{fMB}) in which the gravitational
potential appears explicitly. The Boltzmann distribution can also
be obtained by maximizing the entropy $S$ at fixed mass and energy
or by maximizing the free energy $J$ at fixed mass and
temperature. The gravitational potential is determined
self-consistently by solving the mean field equation
\begin{equation}
\Delta \Phi =4\pi G A e^{-\beta \Phi},
\label{Mf}
\end{equation}
obtained by substituting the density (\ref{fMB}) in the Poisson
equation (\ref{Poisson}). This Boltzmann-Poisson equation has been
studied in relation with the structure of isothermal stellar cores
\cite{chandra} and globular clusters \cite{bt}. It is well-known
that the density of an isothermal gas decreases at large distances
like $r^{-2}$ resulting in the infinite mass problem if the system
is not bounded.

\begin{figure}
\centerline{ \psfig{figure=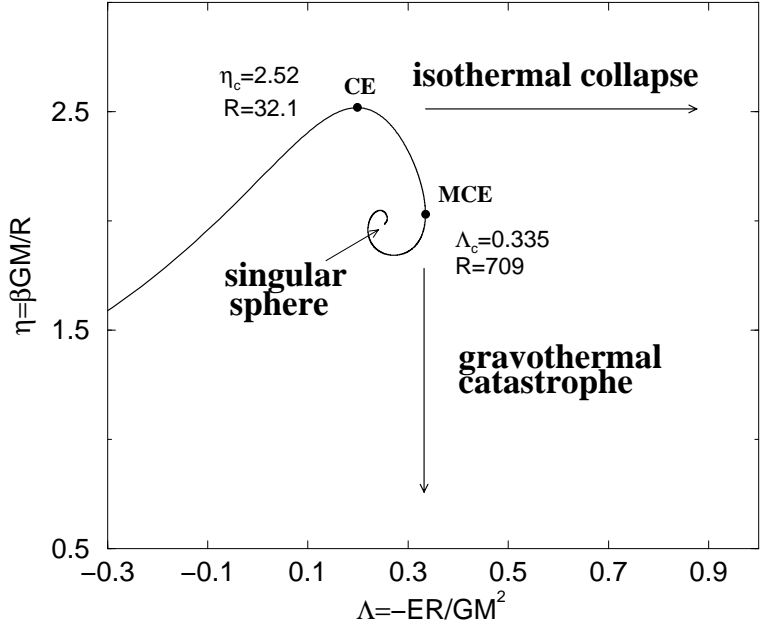,angle=0,height=8.5cm}}
\caption{Equilibrium phase diagram for classical isothermal
spheres. The spiral rolls up indefinitely towards the singular
isothermal sphere $\rho_{s}=1/2\pi G\beta r^{2}$.} \label{pdiag}
\end{figure}

The equilibrium phase diagram $(E,T)$ of isothermal configurations
confined within a box is represented in Fig. \ref{pdiag} where we have
plotted the normalized inverse temperature $\eta={\beta G M/ R}$ as a
function of the normalized energy $\Lambda=-{ER/ GM^{2}}$.  The curve
has a striking spiral behavior parameterized by the density contrast
${\cal R}=\rho(0)/\rho(R)$ going from $1$ (homogeneous system) to
$+\infty$ (singular sphere) as we proceed along the spiral. There is
no equilibrium state above $\Lambda_{c}=0.335$ or $\eta_{c}=2.52$. In
that case, the system is expected to collapse indefinitely. It is also
important to recall that the statistical ensembles are not
interchangeable for systems with long-range interaction, like
gravity. In the microcanonical ensemble, the series of equilibria
becomes unstable after the first turning point of energy $(MCE)$
corresponding to a density contrast of $709$. At that point, the
isothermal spheres pass from local entropy maxima to saddle points. In
the canonical ensemble, the series of equilibria becomes unstable
after the first turning point of temperature $(CE)$ corresponding to a
density contrast of $32.1$. At that point, the isothermal spheres pass
from maxima of free energy to saddle points.  It can be noticed that
the region of negative specific heats between $(CE)$ and $(MCE)$ is
stable in the microcanonical ensemble but unstable in the canonical
ensemble as expected on general physical grounds \cite{pad}. The
thermodynamical stability of isothermal spheres can be deduced from
the topology of the $\beta-E$ curve by using the method of Katz
\cite{katz78} who has extended Poincar\'e's theory of linear series of
equilibria.  The stability problem can also be reduced to the study of
an eigenvalue equation associated with the second order variations of
entropy or free energy as studied by Padmanabhan \cite{pad2} in MCE
and Chavanis
\cite{chavcano} in CE.  The same stability limits as Katz are obtained
but this method provides in addition the form of the density
perturbation profiles that trigger the instability at the critical
points. We also recall that isothermal spheres are at most metastable:
there is no {\it global} maximum of entropy or free energy for a
classical system of point masses in gravitational interaction
\cite{antonov}.

\subsection{Linear stability analysis}
\label{sec_stab}

We now perform a linear stability analysis of the SP system. Let
$\rho$, $T$ and $\Phi$ refer to a stationary solution of
Eq.~(\ref{X5}) and consider a small perturbation $\delta \rho$,
$\delta T$ and $\delta\Phi$ around this solution that does not change
energy and mass. Since a stationary solution of the SP system is a
critical point of entropy, we must assume $\Lambda\le
\Lambda_{c}$ for a solution to exist. Writing $\delta \rho\sim e^{\lambda
t}$ and expanding Eq.~(\ref{X5}) to first order, we find that
\begin{equation}
\lambda\delta \rho={1\over r^{2}}{d\over dr}\biggl \lbrack
{r^{2}\over \xi}\biggl (\delta T {d\rho\over dr}+
T{d\delta \rho\over dr}+\delta \rho {d\Phi\over dr}+
\rho{d\delta\Phi\over dr}\biggr )\biggr\rbrack.
\label{l1}
\end{equation}
It is convenient to introduce the notation
\begin{equation}
\delta \rho={1\over 4\pi r^{2}}{dq\over dr}.
\label{l2}
\end{equation}
Physically, $q$ represents the mass perturbation $q(r)\equiv \delta
M(r)=\int_{0}^{r}4\pi {r'}^{2}\delta \rho(r')dr'$ within the sphere of
radius $r$. It satisfies therefore the boundary conditions
$q(0)=q(R)=0$. Substituting Eq.~(\ref{l2}) in Eq.~(\ref{l1}) and
integrating, we obtain
\begin{equation}
{\lambda\xi\over r^{2}} q= 4\pi \delta T {d\rho\over dr}+T{d\over
dr}\biggl ({1\over r^{2}}{dq\over dr}\biggr )+{1\over r^{2}}{dq\over
dr}{d\Phi\over dr}+4\pi \rho {d\delta\Phi\over dr},
\label{l3}
\end{equation}
where we have used $q(0)=0$ to eliminate the constant of
integration. Using the condition of hydrostatic equilibrium
$T{d\rho/dr}+\rho{d\Phi/dr}=0$ and the  Gauss theorem
${d\delta\Phi/dr}={Gq/r^{2}}$, we can rewrite Eq. (\ref{l3}) as
\begin{equation}
{\lambda\xi\over 4\pi\rho Tr^{2}} q=- {\delta T\over T^{2}}
{d\Phi\over dr}+{1\over 4\pi \rho}{d\over dr}\biggl ({1\over
r^{2}}{dq\over dr}\biggr )-{1\over 4\pi \rho^{2}}{1\over
r^{2}}{dq\over dr}{d\rho\over dr}+{Gq\over Tr^{2}}, \label{l3bis}
\end{equation}
or, alternatively,
\begin{equation}
{d\over dr}\biggl ({1\over 4\pi \rho r^{2}}{dq\over dr}\biggr
)+{Gq\over Tr^{2}}-{\lambda\xi\over 4\pi \rho T r^{2}} q-{\delta
T\over T^{2}} {d\Phi\over dr}=0.
\label{l8bis}
\end{equation}
From the energy constraint (\ref{EE2}) we find that
\begin{equation}
\delta T=-{2\over 3M}\int_{0}^{R}\delta \rho\Phi 4\pi r^{2}\,dr={2\over 3M}
\int_{0}^{R}q {d\Phi\over dr} \,dr.
\label{l9}
\end{equation}
Hence, our linear stability analysis leads to the eigenvalue equation
\begin{equation}
{d\over dr}\biggl ({1\over 4\pi \rho r^{2}}{dq\over dr}\biggr
)+{Gq\over T r^{2}}-{2V\over 3 M T ^{2}} {d\Phi\over
dr}={\lambda\xi\over 4\pi \rho T r^{2}}q,
\label{l11}
\end{equation}
where
\begin{equation}
V=\int_{0}^{R}q {d\Phi\over dr} dr,
\label{l12}
\end{equation}
where we recall that $q(0)=q(R)=0$. Eq.~(\ref{l11}) is similar to
the eigenvalue equation associated with the second order
variations of entropy found by Padmanabhan \cite{pad2}. In
particular, they coincide for marginal stability ($\lambda =0$).
More generally, it is proven in Appendix \ref{sec_connexion2} that
a stationary solution of Eq.~(\ref{X5}) is linearly stable if and
only if it is a local entropy maximum. The zero eigenvalue
equation was solved by Padmanabhan \cite{pad2}.  It is found that
marginal stability occurs at the point of minimum energy
$\Lambda=\Lambda_{c}$, in agreement with Katz \cite{katz78}
approach, and that the perturbation $\delta \rho/\rho$ that
induces instability (technically the eigenfunction associated with
$\lambda=0$) has a ``core-halo'' structure (i.e., two nodes). It
is also argued qualitatively that the number of oscillations in
the profile $\delta \rho/\rho$ increases as we proceed along the
series of equilibria, see Fig.~\ref{pdiag}, up to the singular
sphere (i.e for higher and higher density contrasts). Of course,
on the upper branch of Fig.~\ref{pdiag}, the eigenvalues $\lambda$
are all negative (meaning stability) while more and more
eigenvalues become positive (meaning instability) as we spiral
inward for ${\cal R}>709$.

If we fix the temperature $T$ instead of the energy $E$, the eigenvalue
equation becomes (take $\delta T=0$ in Eq.~(\ref{l8bis})):
\begin{equation}
{d\over dr}\biggl ({1\over 4\pi \rho r^{2}}{dq\over dr}\biggr
)+{Gq\over T r^{2}}={\lambda\xi\over 4\pi \rho T r^{2}} q.
\label{l13}
\end{equation}
This is similar to the equation obtained by Chavanis
\cite{chavcano} by analyzing the second order variations of free
energy. The case of marginal stability ($\lambda=0$) coincides
with the point of minimum temperature $\eta=\eta_{c}$ like in Katz
\cite{katz78} analysis. It is found that the perturbation $\delta
\rho/\rho$ that induces instability at $\eta=\eta_{c}$ in the
canonical ensemble has {\it not} a ``core-halo'' structure (it has
only one node).

\section{Self-similar solutions of the Smoluchowski-Poisson system}
\label{sec_selfsimilar}

\subsection{Formulation of the general problem}
\label{sec_general}

We now describe the collapse regime and look for self-similar
solutions of the SP system. Restricting ourselves to spherically
symmetric solutions and using the Gauss theorem, we obtain the
integrodifferential equation
\begin{equation}
{\partial\rho\over\partial t}={1\over r^{2}}{\partial\over\partial
r}\biggl\lbrace {r^{2}\over\xi}\biggl (T{\partial\rho\over\partial
r}+{1\over r^{2}}G\rho\int_{0}^{r} \rho(r')4\pi r^{'2}dr'\biggr )\biggr\rbrace.
\label{self1bis}
\end{equation}
We look for self-similar solutions in the form
\begin{equation}
\rho(r,t)=\rho_{0}(t)f\biggl ({r\over r_{0}(t)}\biggr ),
\qquad r_{0}=\biggl ({T\over G\rho_{0}}\biggr )^{1/2},
\label{self2}
\end{equation}
where the density $\rho_{0}(t)$ is of the same order as the central
density $\rho(0,t)$ and the radius $r_{0}$ is of the same order as the
King radius $r_{K}=(9T/4\pi G\rho(0))^{1/2}$ which gives a good
estimate of the core radius of a stellar system
\cite{bt}. Substituting the {\it ansatz} (\ref{self2}) into Eq.
(\ref{self1bis}), we find that
\begin{equation}
{d\rho_{0}\over dt}f(x)-{\rho_{0}\over r_{0}}{dr_{0}\over
dt}xf'(x)={G\rho_{0}^{2}\over\xi}{1\over x^{2}}{d\over dx}\biggl
\lbrace x^{2}\biggl (f'(x)+{1\over x^{2}}f(x)\int_{0}^{x}f(x')4\pi
x^{'2}dx'\biggr )\biggr \rbrace,
\label{self4}
\end{equation}
where we have set $x=r/r_{0}$.  The variables of position and time
separate provided that there exists $\alpha$ such that $\rho_{0}
r_{0}^{\alpha}\sim 1$. In that case, Eq.~(\ref{self4}) reduces to
\begin{equation}
{d\rho_{0}\over dt}\biggl (f(x)+{1\over \alpha}xf'(x)\biggr
)={G\rho_{0}^{2}\over\xi}{1\over x^{2}}{d\over dx}\biggl \lbrace
x^{2}\biggl (f'(x)+{1\over x^{2}}f(x)\int_{0}^{x}f(x')4\pi
x^{'2}dx'\biggr )\biggr \rbrace.
\label{self6}
\end{equation}
Assuming that such a scaling exists implies that
$(\xi/G\rho_{0}^{2})(d\rho_{0}/dt)$ is a constant that we arbitrarily
set to be equal to $1$. This leads to
\begin{equation}
\rho_{0}(t)={\xi\over G}(t_{coll}-t)^{-1},
\label{self7}
\end{equation}
so that the central density becomes infinite in a finite time
$t_{coll}$ while the core shrinks to zero as
$r_{0}\sim (t_{coll}-t)^{1/\alpha}$. Since the collapse time appears as an
integration constant, its precise value cannot be explicitly
determined. The scaling equation now reads
\begin{equation}
f(x)+{1\over \alpha}xf'(x)={1\over x^{2}}{d\over dx}\biggl \lbrace
x^{2}\biggl (f'(x)+{1\over x^{2}}f(x)\int_{0}^{x}f(x')4\pi
x^{'2}dx'\biggr )\biggr \rbrace, \label{self8}
\end{equation}
which determines the invariant profile $f(x)$. Alternative forms of
Eq.~(\ref{self8}) are given in Appendix \ref{sec_scalingeq}. If one
knows the value of $\alpha$, Eq.~(\ref{self8}) leads to a ``shooting
problem'' where the value of $f(0)$ is uniquely selected by the
requirement of a reasonable behavior for $f(x)$ at large distances
(see below). As $f(x)\rightarrow 0$ for large $x$, we can only keep
the leading terms in Eq.  (\ref{self8}), which leads to $f(x)\sim
x^{-\alpha}$ when $x\rightarrow +\infty$.

The velocity profile defined by Eq. (\ref{X6}) can be written
\begin{equation}
u(r,t)=-v_{0}(t)V\biggl ({r\over r_{0}(t)}\biggr ),
\label{vel1}
\end{equation}
with
\begin{equation}
v_{0}(t)={T\over \xi r_{0}}\qquad {\rm and} \qquad V(x)={f'(x)\over
f(x)}+{4\pi\over x^{2}}\int_{0}^{x}f(x')x^{'2}\,dx'.
\label{vel2}
\end{equation}
The invariant profile $V(x)$ has the asymptotic behaviors
$V(x)\sim x$ when $x\rightarrow 0$ and $V(x)\sim {1/ x}$ when
$x\rightarrow +\infty$. On the other hand, the mass profile can be
written
\begin{equation}
M(r,t)=M_{0}(t)g\biggl ({r\over r_{0}(t)}\biggr ),
\label{masspro1}
\end{equation}
with
\begin{equation}
M_{0}(t)=\rho_{0}r_{0}^{3}\qquad {\rm and} \qquad
g(x)=4\pi\int_{0}^{x}f(x')x^{'2}dx'. \label{masspro2}
\end{equation}
The invariant profile $g(x)$ has the asymptotic behaviors
$g(x)\sim x^{3}$ when $x\rightarrow 0$ and $g(x)\sim x^{3-\alpha}$
when $x\rightarrow +\infty$.

\subsection{Canonical ensemble}
\label{sec_sscano}

In the canonical ensemble in which the temperature $T$ is a constant,
Eq.~(\ref{self2}) leads to $\alpha=2$ (the particular case $T=0$ is treated
in Appendix \ref{sec_cold}).  In that case, the scaling
equation (\ref{self8}) can be solved analytically (see Appendix
\ref{sec_scalingeq}) and the invariant profile is exactly given by
\begin{equation}
f(x)={1\over 4\pi}{6+x^{2}\over \bigl (1+{x^{2}\over 2}\bigr )^{2}}.
\label{self10}
\end{equation}
This solution satisfies $f(0)={3\over 2\pi}$ and $f(x)\sim {1\over
\pi x^{2}}$ as $x\rightarrow +\infty$. From Eq.~(\ref{self7}), we
find that the central density  and the core radius evolve with
time as
\begin{equation}
\rho(0,t)=\rho_{0}(t)f(0)={3\xi\over 2\pi G}
(t_{coll}-t)^{-1},\qquad  r_{0}(t)=\biggl ({T\over\xi}\biggr
)^{1/2}(t_{coll}-t)^{1/2}. \label{self12}
\end{equation}
On the other hand, using Eq. (\ref{self10}), we find that the velocity
profile and the mass profile are given by Eqs. (\ref{vel1}) and
(\ref{masspro1}) with
\begin{equation}
v_{0}(t)= \biggl ({T\over\xi}\biggr
)^{1/2}(t_{coll}-t)^{-1/2}\qquad {\rm and}\qquad   V(x)={2x\over
6+x^{2}}, \label{fnd2}
\end{equation}
\begin{equation}
M_{0}(t)= \biggl ({T^{3}\over \xi G^{2}}\biggr
)^{1/2}(t_{coll}-t)^{1/2}\qquad {\rm and}\qquad
g(x)={4x^{3}\over 2+x^{2}}. \label{masw2}
\end{equation}
At $t=t_{coll}$, the scaling solutions (\ref{self2}) (\ref{fnd2}) and
(\ref{masw2}) converge to the singular profiles
\begin{equation}
\rho(r,t=t_{coll})={T\over \pi G r^{2}}, \qquad u(r,t=t_{coll})=
-{2T\over\xi r},  \qquad M(r,t=t_{coll})={4T\over G }r.
\label{fnd3}
\end{equation}
It is interesting to note that the density profile (\ref{fnd3}) has
the same $r$-dependence as that of the singular solution to the static
isothermal gas sphere $\rho=1/2\pi G\beta r^{2}$ \cite{bt}, the two
profiles just differing by a factor of $2$. Therefore, the
relationship between the density and the gravitational potential in
the tail of the scaling profile is given by a Boltzmann distribution
\begin{equation}
\rho\sim A e^{-{1\over 2T}\Phi},
\label{fnrr}
\end{equation}
with a temperature $2T$ instead of $T$. A $r^{-2}$ decay of the
density at large distances was also found by Penston
\cite{penston} in his investigation of the self-similar collapse
of isothermal gas spheres described by the Euler equations. This
is a general characteristic of the collapse in the canonical
ensemble ($T={\it Cst.}$). It should be noticed that the free
energy does not diverge at $t_{coll}$ although the system
undergoes a complete collapse. Therefore, at $t=t_{coll}$, the
density profile is {\it not} a Dirac peak contrary to what might
be expected from rigorous results of statistical mechanics
\cite{kiessling}. In fact, there is no contradiction because the
Dirac peak is formed during the post collapse evolution
\cite{scgravite}.

We now show that the self-similar solution (\ref{self2}) is not
sufficient to quantitatively describe the full density profile
(especially when $r\sim R$). To understand the problem, let us
calculate the mass contained in the scaling profile at
$t=t_{coll}$. Using Eq. (\ref{fnd3}), we have
\begin{equation}
M_{scaling}=\int_{0}^{R} {T\over \pi G r^{2}}4\pi r^{2}dr={4R\over G\beta}.
\label{self13}
\end{equation}
The mass $M_{scaling}$ is finite but, in general, it is not equal to
the total mass $M$ imposed by the initial condition. This means that
there must be a non-scaling contribution to the density which should
contain the remaining mass (possibly negative when $M<M_{scaling}$,
i.e. $\eta<4$). That the scaling solution (\ref{self2}) is not an
exact solution of our problem is also visible from the boundary
conditions. Indeed, according to Eq. (\ref{boundary1}) we should
have
\begin{equation}
{\partial\ln\rho\over\partial r}=-{\beta GM\over R^{2}}, \qquad {\rm
for }\qquad r=R.
\label{self15}
\end{equation}
This relation is clearly not satisfied by Eq. (\ref{fnd3}) except
for the particular value $\eta=2$. These problems originate
because we work in a finite container. The scaling solution
(\ref{self2}) would be exact in an infinite domain but, in that case,
the total mass of the system is infinite. In addition, if we remove
the box, the isothermal spheres are always unstable and the
interesting bifurcations between equilibrium and collapsing
states are lost.

Strictly speaking, we expect that the self-similar solution
(\ref{self2}) will describe the density behavior in the scaling limit
defined by
\begin{equation}
t\rightarrow t_{coll}\quad {\rm or} \quad r_{0}\rightarrow 0, \quad
{\rm and}\quad x=r/r_{0} \quad {\rm fixed}.
\label{self16}
\end{equation}
For the reasons indicated above, it probably does not reproduce the
density near the edge of the box, that is for $r\sim R\gg
r_{0}$. Therefore, we write another equation for the density, making
the following {\it ansatz}:
\begin{equation}
\rho(r,t)=\rho_{0}(t)f\biggl ({r\over r_{0}(t)}\biggr )+
{T\over 4\pi G}F(r,t),
\label{self17}
\end{equation}
where $F(r,t)$ is the profile that contains the excess or deficit of
mass. For $t=t_{coll}$, we have
\begin{equation}
\rho(r,t_{coll})={T\over 4\pi G}\biggl ({4\over r^{2}}+F(r)\biggr ),
\label{an1}
\end{equation}
and it would be desirable to find an approximate expression for the
function $F(r)=F(r,t_{coll})$. A differential equation for $F(r)$ can
be obtained by substituting the {\it ansatz} (\ref{self17}) in the
dynamical equation (\ref{self1bis}) and taking the limit
$t=t_{coll}$. We need first to discuss the term $\partial\rho/\partial
t (r,t_{coll})$.  For $t\rightarrow t_{coll}$, we can use the
expansion of the function $f(x)$, given by Eq. (\ref{self10}), to
second order in $1/x^{2}$ to get
\begin{equation}
\rho(r,t)={\rho_{0}r_{0}^{2}\over \pi r^{2}}\biggl (1+{2r_{0}^{2}
\over r^{2}}+...\biggr )+{T\over 4\pi G}F(r,t).
\label{self18}
\end{equation}
Then, using Eqs.~(\ref{self2}) and (\ref{self7}), we obtain to first
order in $t_{coll}-t$:
\begin{equation}
\rho(r,t)=\rho(r,t_{coll})+{T^{2}\over 4\pi G\xi}\biggl \lbrack
{8\over r^{4}}-{\xi\over T}{\partial F\over\partial t}(r,t_{coll})
\biggr \rbrack (t_{coll}-t)+...
\label{self19}
\end{equation}
leading to
\begin{equation}
{\partial \rho\over \partial t}(r,t_{coll})={T^{2}\over 4\pi
G\xi}\biggl \lbrack -{8\over r^{4}}+{\xi\over T}{\partial
F\over\partial t}(r,t_{coll})\biggr \rbrack.
\label{fhn}
\end{equation}
The trouble is that we do not know the function ${\partial
F/\partial t}(r,t_{coll})$. It is possible, however, to derive an exact
integral equation that it must satisfy. Since the exact profile
$\rho(r,t)$ conserves mass, we have just before $t_{coll}$:
\begin{equation}
\int_{0}^{R}{\partial\rho\over\partial t}(r,t_{coll}^{-})r^{2}dr=0.
\label{clem1}
\end{equation}
The scaling profile $\rho_{scaling}(r,t)$ is an exact solution of
Eq. (\ref{self1bis}) but it does not conserve
mass. Multiplying Eq. (\ref{self1bis}) by $r^{2}$ and integrating
from $r=0$ to $R$, we get
\begin{equation}
\int_{0}^{R}{\partial\rho_{scaling}\over\partial t}(r,t_{coll}^{-})r^{2}dr=
{R^{2}\over \xi}\biggl (T{\partial\rho_{scaling}\over\partial r}(R)+
\rho_{scaling}{GM_{scaling}\over R^{2}}\biggr )_{|t=t_{coll}}=
{2T^{2}\over \pi G\xi R},
\label{clem2}
\end{equation}
where we have used Eqs.~(\ref{fnd3}) and (\ref{self13}) to obtain
the last equality. Now, subtracting Eqs.~(\ref{clem1}) and
(\ref{clem2}), using Eq.~(\ref{self17}) and passing to the limit
$t\rightarrow t_{coll}$, we find that
\begin{equation}
\int_{0}^{R}{\partial F\over\partial t}(r,t_{coll})r^{2}dr=-{8T\over \xi R}.
\label{clem3}
\end{equation}
This relation implies in particular that we cannot take $(\partial
F/\partial t)(r,t_{coll})=0$ in Eq.~(\ref{fhn}). In fact, it is likely
that $F(r,t)$ involves combinations of the type
\begin{equation}
F(r,t)\sim \rho_{0}f(r/r_{0})r^{2}F(r),\ {1\over
r^{2}}(r^{2}+cr_{0}^{2})F(r),\ F(\sqrt{r^{2}+c r_{0}^{2}}),...
\label{clem4}
\end{equation}
which reduce to $F(r)$ in the limit $t\rightarrow
t_{coll}$. Considering the time derivative of these expressions at
$t=t_{coll}$, we find that they take only one of the two forms
$F(r)/r^{2}$ and $F'(r)/r$. We are therefore led to make the following
{\it ansatz}~:
\begin{equation}
{\xi\over T}{\partial F\over\partial t}(r,t_{coll})=a{F(r)\over
r^{2}}+b{F'(r)\over r},
\label{clem7}
\end{equation}
where $a$ and $b$ are some unknown constants which will be determined
by an optimization procedure (see below). If we substitute the {\it
ansatz} (\ref{self17}) in Eq.~(\ref{self1bis}), take the limit
$t=t_{coll}$ and use Eqs.~(\ref{fhn}) and (\ref{clem7}), we find after
some simplifications that $F(r)$ satisfies the differential equation
\begin{equation}
r^{2}F''+(6-b)r{F'}+r^{2}F^{2}+(8-a)F+F'\int_{0}^{r}F(x)x^{2}dx-{8\over
r^{3}}\int_{0}^{r}F(x)x^{2}dx=0.
\label{self20}
\end{equation}
Interestingly, the final profile equation (\ref{self20}) is {\it not}
obtained by setting $\partial\rho/\partial t=0$ in the dynamical
equation as, even in the stationary looking tail,
$\partial\rho/\partial t$ is in fact of order $1$ due to the fast
collapse dynamics.

Equation (\ref{self20}) leads to another ``shooting problem'',
starting this time from $r=R$. The value $F(R)$ is selected by
imposing the condition that the total mass is $M$. This yields
\begin{equation}
\int_{0}^{R}F(r)r^{2}dr=\beta G \biggl (M-{4R\over  \beta G}\biggr ),
\label{self22}
\end{equation}
where $4R/ \beta G=M_{scaling}$ is the mass included in the scaling
part. Moreover, $F'(R)$ is fully determined by the boundary condition
(\ref{boundary1}) at $r=R$ which implies, together with
Eq. (\ref{an1}),
\begin{equation}
F'(R)+{\beta GM\over R^{2}}F(R)={8\over R^{3}}-{4\beta GM\over R^{4}}.
\label{self23}
\end{equation}
Finally, the exact relation (\ref{clem3}) combined with Eq. (\ref{clem7})
imposes the condition
\begin{equation}
(a-b)\int_{0}^{R}F(r)\,dr+bRF(R)=-{8\over R}.
\label{clem8}
\end{equation}
In order to determine the values of $a$ and $b$ we shall require that the
value of the total density at $r=R$ is maximum, as the system would
certainly tend to expel some mass if it were not bound to a sphere
(recall that the profile $F$ arises because of boundary effects). In
addition, Eq. (\ref{clem8}) implies that $F$ is integrable, so
that the optimization process should be performed including this
constraint (if $F$ is integrable, then Eq.~(\ref{clem8}) is
automatically satisfied as it is equivalent to the conservation of
mass). In the section devoted to numerical simulations, we study $F$
numerically and compare it with the numerical profiles obtained by solving the SP system.

\subsection{Microcanonical ensemble}
\label{sec_ssmicro}

If the temperature is not fixed but determined by the energy
constraint (\ref{EE2}), then the exponent $\alpha$ is not known {\it a
priori}. However, we have solved Eq.~(\ref{self8}) numerically for
different values of $\alpha$ and found that there is a maximum value
for $\alpha$ above which Eq.~(\ref{self8}) does not have any physical
solution. This value $\alpha_{max}=2.20973304...$ is close to that
found by Lynden-Bell \& Eggleton \cite{lbe} (and, to some extent, by
Cohn \cite{cohn} and Larson \cite{larson}) in their investigation on
the gravitational collapse of globular clusters. The common point
between these models is that the temperature is free to diverge so the
scaling exponent $\alpha$ cannot be determined from simple dimensional
analysis. However, the agreement on the value of $\alpha$ is probably
coincidental since our model differs from the others in many respects.

In the present case, $\alpha_{max}$ is just an upper bound on $\alpha$
not a unique eigenvalue determined by the scaling equations like in
Ref. \cite{lbe} for example. However, this maximum value leads to the
fastest divergence of the entropy and the temperature so it is
expected to be selected by the dynamics (recall that the SP system is
consistent with a maximum entropy production principle
\cite{csr}). Indeed, the temperature and the entropy respectively
diverge like
\begin{equation}
T(t)\sim (t_{coll}-t)^{-{\alpha-2\over\alpha}}, \qquad S(t)\sim
-{3(\alpha-2)\over 2\alpha}\ln (t_{coll}-t).
\label{self25}
\end{equation}
Note that these divergences are quite weak as the exponent
involved is small
${\alpha_{max}-2\over\alpha_{max}}=0.0949133...$. For
$\alpha=\alpha_{max}$, the value of $f(0)$ selected by the
shooting problem defined by Eq. (\ref{self8}) is $f(0)=5.178...$.
Therefore, the central density evolves with time as
\begin{equation}
\rho(0,t)=5.178...{\xi\over G} (t_{coll}-t)^{-1}.
\label{self27}
\end{equation}
The coefficient in front of $(t_{coll}-t)^{-1}$ is approximately
$10$ times larger than for $\alpha=2$ (see Eq.~(\ref{self12})).
The density profile at $t=t_{coll}$ is equal to
\begin{equation}
\rho(r,t=t_{coll})= {K\over r^{\alpha}},
\label{fl2}
\end{equation}
where $K$ is a constant which is not determined by the scaling theory.
Using Eq.~(\ref{fl2}) and the Gauss theorem, we find
that the relation between $\rho$ and $\Phi$ in the tail of the
self-similar profile is that of a {\it polytrope}:
\begin{equation}
\rho\sim (\Phi-{\it Cst.})^{\alpha\over\alpha-2},
\label{fl3}
\end{equation}
with index $n=\alpha/(\alpha-2)\simeq 10.53$ for $\alpha=\alpha_{max}$.

We now address the divergence of the potential energy which should
match that of the temperature (or kinetic energy) in order to ensure
energy conservation. After an integration by parts, the potential
energy can be written
\begin{equation}
W=-{GM^{2}\over 2R}-{1\over 8\pi G}\int (\nabla\Phi)^{2}\,d^{3}{\bf r}.
\label{self30}
\end{equation}
Then, using the Gauss theorem, we obtain
\begin{equation}
W=-{GM^{2}\over 2R}-{G\over 2}\int_{0}^{R}{1\over r^{2}}\biggl
(\int_{0}^{r}\rho(r')4\pi r^{'2}\,dr'\biggr )^{2}\,dr.
\label{self31}
\end{equation}
If we assume that all the potential energy is in the scaling profile,
we get a contradiction since
\begin{equation}
W_{scaling}(t=t_{coll})\sim -{G\over 2}\int_{0}^{R}{1\over
r^{2}}\biggl (\int_{0}^{r} {1\over r^{'\alpha}} 4\pi r^{'2}\,dr'\biggr
)^{2}dr\sim -\int_{0}^{R} r^{4-2\alpha}\,dr,
\label{self32}
\end{equation}
converges for $\alpha<5/2$. Since the temperature diverges with time
for $\alpha=\alpha_{max}$, the total energy cannot be conserved. This
would suggest that $\alpha=2$ like in the canonical ensemble. We
cannot rigorously exclude this possibility but a value of $\alpha$
close to $\alpha_{max}\simeq 2.21$ is more consistent with the
numerical simulations (see Sec. \ref{sec_simulations}) and leads to a
larger increase of entropy (in agreement with the MEPP). If this value
is correct, the divergence of the gravitational energy should
originate from the non scaling part of the profile which also
accommodates for the mass conservation. In the following, a possible
scenario allowing for the gravitational energy to diverge is
presented.

Let us assume that there exists two length scales $r_{1}$ and
$r_{2}$ satisfying $r_{0}\ll r_{1}\ll r_{2}\ll R$ with
$r_{0},r_{1},r_{2}\rightarrow 0$ for $t\rightarrow t_{coll}$ such
that the mass between $r_{1}$ and $r_{2}$ is of order $1$. The
physical picture that we have in mind is that this mass will
progress towards the center of the domain and form a dense nucleus
with larger and larger potential energy. We assume that for
$r_{1}<r<r_{2}$ the density behaves as
\begin{equation}
\rho(r,t)\sim {r_{1}^{\gamma-\alpha}\over r^{\gamma}},
\label{self34}
\end{equation}
so that this functional form matches with the scaling profile for
$r\sim r_{1}$. If we impose that the total mass between $r_{1}$ and
$r_{2}$ is of order $1$, we get
\begin{equation}
\int_{r_{1}}^{r_{2}} {r_{1}^{\gamma-\alpha}\over r^{\gamma}} r^{2}dr\sim
1, \qquad {\rm i.e.} \qquad r_{2}\sim r_{1}^{\alpha-\gamma\over 3-\gamma},
\label{self35}
\end{equation}
which shows that $r_{2}\gg r_{1}$ since $\alpha<3$. Now, the
contribution to the potential energy of the density between $r_{1}$
and $r_{2}$ which is assumed to be the dominant part is
\begin{equation}
W\sim -\int_{r_{1}}^{r_{2}}{1\over r^{2}}\biggl
(\int_{r_{1}}^{r}{r_{1}^{\gamma-\alpha}\over
r^{'\gamma}}r^{'2}\,dr'\biggr )^{2}\,dr\sim
-r_{1}^{2(\gamma-\alpha)}r_{2}^{5-2\gamma}\sim
-r_{1}^{-(\alpha-\gamma)/(3-\gamma)},
\label{self36}
\end{equation}
where we have used Eq.~(\ref{self35}) to get the last equivalent.
Since the divergence of the potential energy must compensate that
of the kinetic term we must have $-W\sim {3\over 2}M T \sim
r_{0}^{2-\alpha}$  where we have used Eqs.~(\ref{self2}) to get
the last equivalent. This relation implies that $r_{0}$ and
$r_{1}$ are related to each other by
\begin{equation}
r_{1}\sim r_{0}^{(\alpha-2)(3-\gamma)/(\alpha-\gamma)}.
\label{self38}
\end{equation}
Now, imposing $r_{1}\gg r_{0}$ leads to $\gamma <2$.
Therefore, any value of $\gamma<2$ leads to the correct divergence of
$W$ within this scenario.  Note that Eq.~(\ref{self34}) may arise from
the next correction to scaling of the form
\begin{equation}
\rho(r,t)=\rho_{0}f(r/r_{0})+\rho_{0}^{\overline{\gamma}}f_{1}(r/r_{0})+...,
\label{self40}
\end{equation}
with $f_{1}(x)\sim x^{-\gamma}$ for large $x$ and
$\overline{\gamma}<1$ for the first term to be dominant in the scaling
regime. Matching the large $x$ behavior of
Eqs.~(\ref{self34}) and (\ref{self40}), we obtain
\begin{equation}
\rho_{0}^{\overline{\gamma}}r_{0}^{\gamma}\sim r_{1}^{\gamma-\alpha},
\label{self41}
\end{equation}
which is equivalent to
\begin{equation}
r_{1}\sim r_{0}^{\alpha\overline{\gamma}-\gamma\over \alpha-\gamma}.
\label{self42}
\end{equation}
Since $\overline{\gamma}<1$, this implies that $r_{1}\gg r_{0}$, as
expected. More precisely, comparing with Eq.~(\ref{self38}), we have
\begin{equation}
\overline{\gamma}={\gamma+(\alpha-2)(3-\gamma)\over\alpha},
\label{self43}
\end{equation}
and we check that the condition $\gamma<2$ is equivalent to
$\overline{\gamma}<1$.

\subsection{Analogy with critical phenomena}
\label{sec_critphen}

In this section, we determine the domain of validity of the
scaling regime by using an analogy with the theory of critical
phenomena. For simplicity, we work in the canonical ensemble but we
expect to get similar results in the microcanonical ensemble. For
$\eta=\theta^{-1}=\beta GM/R$ close to $\eta_{c}$, we define
\begin{equation}
\epsilon={|\eta-\eta_{c}|\over\eta_{c}}\sim
{|\theta_{c}-\theta|\over\theta_{c}}\ll 1.
\label{crit0}
\end{equation}

For $\eta=\eta_{c}$ the central density $\rho(0,t)$ goes to a
finite constant $\rho_{\infty}$ when $t\rightarrow +\infty$.  Writing
$\delta\rho(t)=\rho_{\infty}-\rho(0,t)$ and using Eq.~(\ref{X5}),
which is quadratic in $\rho$, we argue that, for $\eta\le \eta_{c}$,
$\delta\rho(t)$ satisfies an equation of the form
\begin{equation}
{d\delta\rho\over dt}\sim {\delta\rho\over\tau}-{G\over\xi}\delta\rho^{2},
\label{crit1}
\end{equation}
where $\tau$ plays the role of a correlation time which is expected to
diverge for $\eta=\eta_{c}$ leading to a slow (algebraic) convergence
of $\delta\rho$ towards $0$ at the critical temperature. Actually, for
$\eta=\eta_{c}$, Eq.~(\ref{crit1}) yields
\begin{equation}
\delta\rho\sim t^{-1}.
\label{crit2}
\end{equation}
Now, if we stand slightly above the critical
point ($\eta>\eta_{c}$), we expect this behavior to hold up to a time
of order $t_{coll}$ for which the perturbation term proportional to
$(1/\xi)(T-T_{c})\Delta \rho(0,t)\sim -\epsilon$ is of the same order
as $\partial\rho/\partial t\sim -1/t^{2}$. This yields
\begin{equation}
t_{coll}\sim \epsilon^{-1/2}\sim (\eta-\eta_{c})^{-1/2}.
\label{crit3}
\end{equation}
By analogy with critical phenomena, it is natural to expect that
$\tau$ has the same behavior for $\eta<\eta_{c}$:
\begin{equation}
\tau \sim (\eta_{c}-\eta)^{-1/2}.
\label{crit4}
\end{equation}
Therefore, for $\eta<\eta_{c}$ and according to Eq.~(\ref{crit1}),
$\delta\rho(t)$ tends exponentially rapidly to the equilibrium value
\begin{equation}
\rho_{\infty}-\rho(0,t=+\infty)={\xi\over G}\ \tau^{-1}\sim
(\eta_{c}-\eta)^{1/2}.
\label{crit4bis}
\end{equation}
This relation is consistent with the results obtained in the
equilibrium study \cite{chavcano}, where the exact result
\begin{equation}
1-\frac{\rho(0)}{\rho_\infty}\approx\left[
\frac{8}{\eta_c-2}\left(1-\frac{\eta}{\eta_c}\right)\right]^{1/2},
\label{stattau}
\end{equation}
is derived close to the critical point.

Another interesting question concerns the extent of the scaling regime
which we expect to be valid for $t_{coll}-t < \delta
t\sim\epsilon^{\nu}$. To compute $\nu$, we integrate the dynamical
equation in the regime where the perturbation $-\epsilon\Delta\rho$
dominates:
\begin{equation}
{\partial\rho\over\partial t}\simeq -\epsilon\Delta\rho,
\label{crit5}
\end{equation}
leading to
\begin{equation}
\rho(0,t)\sim \int_{k<r_{0}^{-1}}k^{2}{\rm exp}(k^{2}\epsilon t)dk\sim
r_{0}^{-2}{\rm exp}(r_{0}^{-2}\epsilon t),
\label{crit6}
\end{equation}
where we have introduced an upper momentum cut-off of order
$r_{0}^{-1}$ to prevent the integral from diverging.  Indeed, the
Laplacian of $\rho$ should become positive for $r\gg r_{0}$ as
$\Delta(r^{-2})=2r^{-4}>0$. Thus, for $\epsilon\ll 1$, we expect
that the density will first saturate to $\rho_{\infty}$ for a long
time of order $t_{coll}$ [see Eq. (\ref{crit2})], before rapidly
increasing [see Eq. (\ref{crit6})], and ultimately reaching the
scaling regime [see Eq. (\ref{self2})]. Comparing
Eq.~(\ref{crit6}) with the density in the scaling regime
$\rho(0,t)\sim r_{0}^{-2}$, we find that the scaling regime is
reached at a time $t_{*}$ such that $r_{0}^{-2}\epsilon t_{*}\sim
1$ (for the argument in the exponential to be of order $1$). Since
$r_{0}\sim (t_{coll}-t)^{1/2}$ in the scaling regime, we get $
t_{coll}-t_{*} \sim \epsilon t_{coll} $. Therefore, the width of
the scaling regime, $\delta t=t_{coll}-t_{*}$, behaves like
\begin{equation}
\delta t\sim t_{coll}\epsilon\sim \epsilon^{1/2},
\label{crit7}
\end{equation}
establishing $\nu=1/2$. Close to the critical point, the collapse
occurs at a very late time and the width of the scaling regime is very
small. Therefore, if we are close to the critical point, it will be
difficult to reach numerically the regime in which the results of
sections \ref{sec_general}-\ref{sec_ssmicro} are valid.

Regrouping all these results, and using again an analogy with critical
phenomena, we expect that the central density obeys the following
equation
\begin{equation}
\rho(0,t)=(t_{coll}-t)^{-1}G(t_{coll}(t_{coll}-t)),
\label{crit8}
\end{equation}
where $t_{coll}\sim \epsilon^{-1/2}$ and the scaling function $G$ satisfies
\begin{equation}
G(0)={3\over 2\pi}, \qquad G(x)\sim {\rho_{\infty}\sqrt{x}}, \quad {\rm
for}\ x\rightarrow +\infty.
\label{crit9}
\end{equation}

\section{Numerical simulations}
\label{sec_simulations}

In this section, we perform direct numerical simulations of the
SP system and compare the results of the simulations
with the theoretical results of Secs. \ref{sec_brown} and
\ref{sec_selfsimilar}. In most of numerical experiments, we start from a
homogeneous sphere with radius $R$ and density $\rho_{*}={3M/4\pi
R^{3}}$.  This configuration has a potential energy $W_{0}=-{3GM^{2}/
5R}$. In the canonical ensemble the temperature is equal to $T$ at any
time. In the microcanonical ensemble, the initial temperature $T_{0}$
is adjusted in order to have the desired value of
$\Lambda={3/5}-{3RT_{0}/ 2GM}$.  By changing the temperature or the
energy, we can explore the whole bifurcation diagram in parameter
space and check the theoretical predictions of
Secs. \ref{sec_brown} and \ref{sec_selfsimilar}.  In the numerical work,
we use dimensionless variables so that $M=R=G=\xi=1$.

\subsection{Microcanonical ensemble}
\label{sec_micro}

We first solve the SP system with the constraint (\ref{EE2}) insuring
the conservation of energy. We confirm the predictions of the
thermodynamical approach in the microcanonical ensemble.  For
$\Lambda=0.334<\Lambda_{c}$, the quantities $\rho(0,t)$, $T(t)$,
$r_{K}(t)$ and $S(t)$ converge to finite values and the system settles
down to a stable thermodynamical equilibrium state with a density contrast
${\cal R}\simeq 596$ less than the critical value $\sim 709$ found by
Antonov \cite{antonov}. At large distances, the density decays
approximately as $r^{-2}$ like the singular isothermal sphere
\cite{chandra}. For $\Lambda=0.359>\Lambda_{c}$, the behavior of
the system is completely different: $\rho(0,t)$ and $T(t)$ diverge to
$+\infty$ and $r_{K}(t)$ goes to zero in a finite time $t_{coll}$. We
were able to follow this ``gravothermal catastrophe'' up to a density
contrast ${\cal R}\sim 10^{4}$. The entropy ${S}(t)$ also diverges to
$+\infty$, but its evolution is slower (logarithmic).  For
$\Lambda=0.335=\Lambda_{c}^{+}$, the system first tends to converge
towards an equilibrium state but eventually collapses.

\begin{figure}[htbp]
\centerline{ \psfig{figure= 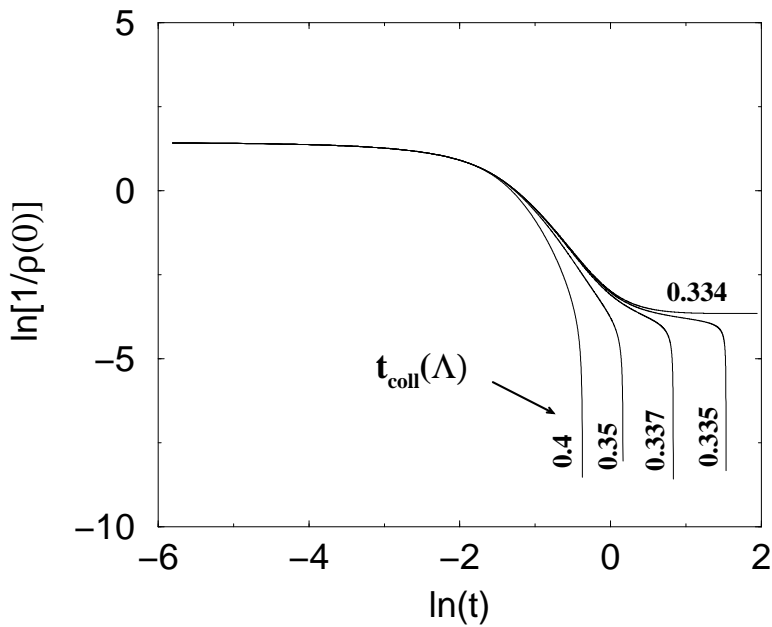,angle=-90,height=8.5cm}}
\caption{Time evolution of the central density for different
values of $\Lambda$. The central density $\rho(0,t)$ becomes
infinite in a finite time $t_{coll}(\Lambda)$ depending on the
value of energy $\Lambda$ (labeling the curves). The figure shows
that the collapse time diverges as we approach the critical value
$\Lambda_{c}=0.3345$ for which a local entropy maximum exists. }
\label{ntlambda}
\end{figure}

\begin{figure}[htbp]
\centerline{ \psfig{figure=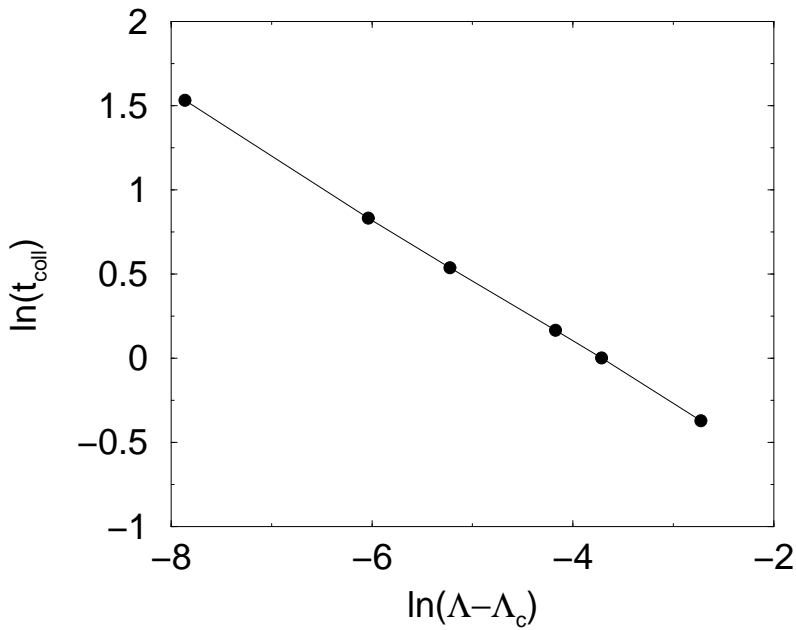,angle=-90,height=8.5cm}}
\caption{Evolution of the collapse time $t_{coll}$ with $\Lambda$.
The figure displays a scaling law $t_{coll}\sim
(\Lambda-\Lambda_{c})^{-\delta}$ with $\delta\simeq 0.4$ close to
the theoretical value $1/2$. } \label{tcoll}
\end{figure}

In Fig.~\ref{ntlambda}, we plot the inverse of the central density as a
function of time for different values of $\Lambda$. For short times,
the density is approximately uniform, as it is initially. In that
case, the diffusion term in Eq.~(\ref{X5}) is negligible and the
system evolves under the influence of the gravitational term
alone. Using the Poisson equation (\ref{Poisson}), the Smoluchowski
equation (\ref{X5}) reduces to
\begin{equation}
{d\rho\over dt}={4\pi G\over\xi}\rho^{2}.
\label{dfnl1}
\end{equation}
Solving for $\rho(t)$, we get
\begin{equation}
\rho(0,t)=\rho_{*} \biggl (1+{4\pi G\over\xi}\rho_{*}
\ t+...\biggr )\qquad (t\rightarrow 0),
\label{dfnl2}
\end{equation}
where $\rho_{*}$ is the initial density. Over longer time scales, a
pressure gradient develops and the two terms in the right hand side of
Eq.~(\ref{X5}) must be taken into account. The system first reaches
a plateau with density $\sim \rho_{\infty}$ (corresponding to an
approximate balance between pressure and gravity) before gravitational
collapse takes place eventually at $t\sim t_{coll}$. In
Fig.~\ref{ntlambda}, we see that the collapse time $t_{coll}$
depends on the value of $\Lambda$ and increases as we approach
the critical value $\Lambda_{c}$. To be more quantitative, we plot in
Fig.~\ref{tcoll} the collapse time $t_{coll}$ as a function of the
distance to the critical point $\Lambda-\Lambda_{c}$. A scaling law is
observed with an exponent $\sim -0.4$ close to the predicted value
$-1/2$ (see Sec. \ref{sec_critphen}).

\begin{figure}[htbp]
\centerline{
\psfig{figure=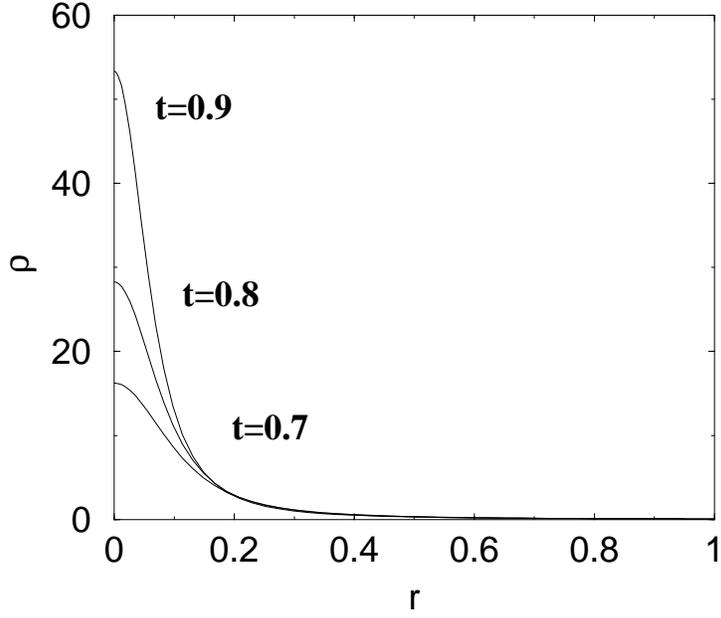,angle=-90,height=8.5cm}}
\caption{Evolution of the density profile for $\Lambda=0.359>\Lambda_{c}$
at different times. Starting from a uniform distribution at $t=0$, the
system develops a ``core-halo'' structure with a shrinking core. From
this figure, we may suspect that the evolution is self-similar,
i.e. the density profiles at different times can be superimposed by an
appropriate rescaling.}
\label{proft}
\end{figure}

\begin{figure}[htbp]
\centerline{
\psfig{figure=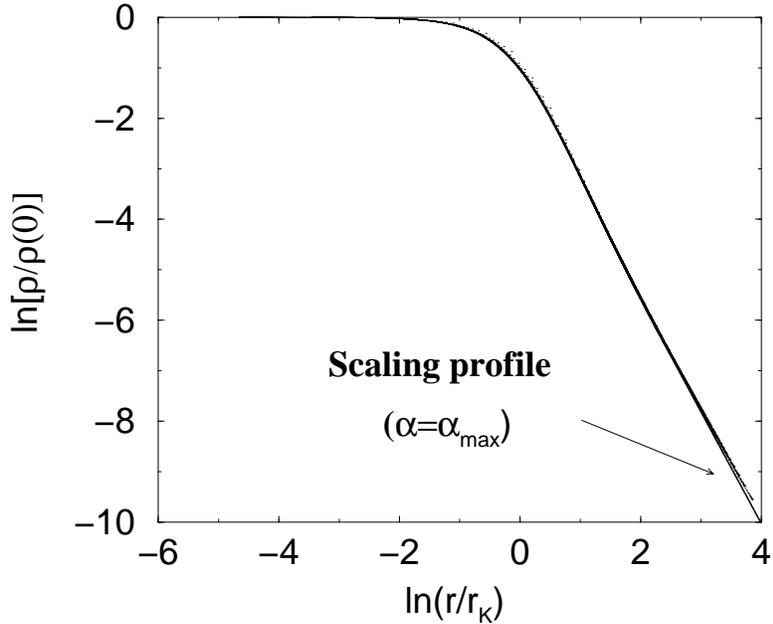,angle=-90,height=8.5cm}}
\caption{This figure represents the (quasi) invariant density
profile obtained for $\Lambda=0.359$ by rescaling the density by the
central density and the radius by the King radius. It is compared with
the theoretical profile $f(x)$ calculated by solving Eq. (\ref{self8}) with
$\alpha=\alpha_{max}$.  }
\label{autosim0.359}
\end{figure}

\begin{figure}[htbp]
\centerline{
\psfig{figure=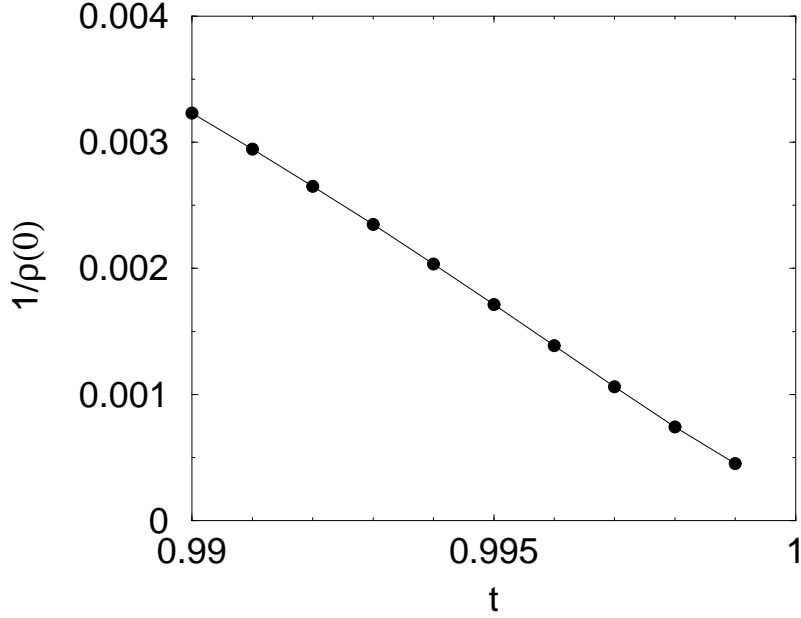,angle=-90,height=8.5cm}}
\caption{Time evolution of the inverse central density for
$\Lambda=0.359$. This curve displays a scaling regime
$1/\rho(0,t)=A(t_{coll}-t)$. The slope of the curve $A\simeq -0.313$
is of the same order as the theoretical value $-1/5.178=-0.193$
corresponding to $\alpha=\alpha_{max}$. The small deviation is
attributed to non scaling corrections.  }
\label{lambda0.359tcoll}
\end{figure}

\begin{figure}[htbp]
\centerline{
\psfig{figure=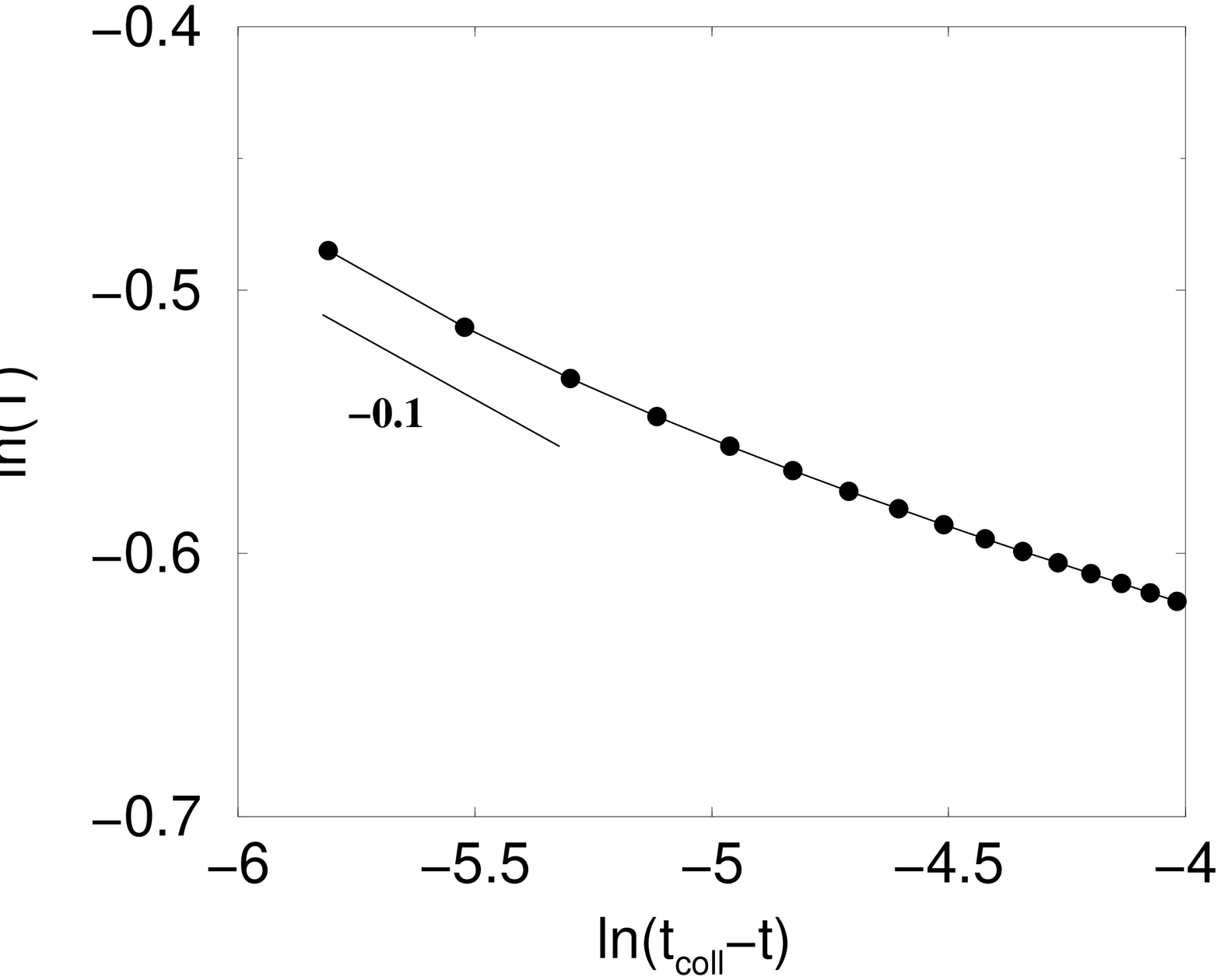,angle=-90,height=8.5cm}}
\caption{Time evolution of the temperature for $\Lambda=0.359$. The curve
displays a scaling regime $T\sim (t_{coll}-t)^{-\gamma}$. The value of
$\gamma\simeq 0.1$ is in agreement with the theoretical value (\ref{self25})
for $\alpha=\alpha_{max}$. }
\label{lambda0.359lnT}
\end{figure}

During the late stage of the collapse, the density profiles are
self-similar that is, they differ only in normalization and scale
(Fig.~\ref{proft}). Indeed, if we rescale the density by the
central density and the radius by the King radius, the density
profiles at various times fall on to the same curve
(Fig.~\ref{autosim0.359}). The invariant profile is compared with
the scaling profile $f(x)$ corresponding to $\alpha=\alpha_{max}$
and the agreement is excellent, except in the tail. This small
discrepancy can be ascribed to the next correction to scaling (see
section \ref{sec_ssmicro}) which generates a power law profile
between $r_{1}$ and $r_{2}$ with an index $\gamma<2$. We have
checked that the logarithmic slope of the profile at $r=R$ is
equal to $-\eta$ in agreement with the boundary condition
(\ref{self15}). However, this relation only holds in a tiny
portion of the curve (invisible in Fig.~\ref{autosim0.359}) so
that the ``effective slope'' is more consistent with a value
$\alpha\simeq 2.2$. In Fig.~\ref{lambda0.359tcoll}, we plot the
inverse central density as a function of time. It is seen that,
for $t\rightarrow t_{coll}$, the central density diverges with
time like $(t_{coll}-t)^{-1}$ in good agreement with the
theoretical expectation. The slope of the curve in Fig.
\ref{lambda0.359tcoll} is approximately $-0.313$ but is
consistently getting closer to the theoretical value
$-1/5.178...\approx -0.193$ corresponding to $\alpha=\alpha_{max}$
as $\Lambda$ increases, or as $t$ approaches $t_{coll}$ (the small
difference is attributed to non scaling corrections, as discussed
in Sec. \ref{sec_ssmicro}). Note that a value of $\alpha=2$ would
yield a much larger slope $-2\pi/3\approx -2.094$ (see Sec.
\ref{sec_sscano}), which is clearly not observed here. Therefore,
the simulations are consistent with a value of
$\alpha=\alpha_{max}$, as expected on physical grounds. This value
$\alpha=\alpha_{max}$ is also consistent with the slow but
existing divergence of the temperature. Indeed, the slope of the
curve in Fig.~\ref{lambda0.359lnT} is approximately $-0.1$ in
agreement with the theoretical expectation.

\begin{figure}[htbp]
\centerline{
\psfig{figure=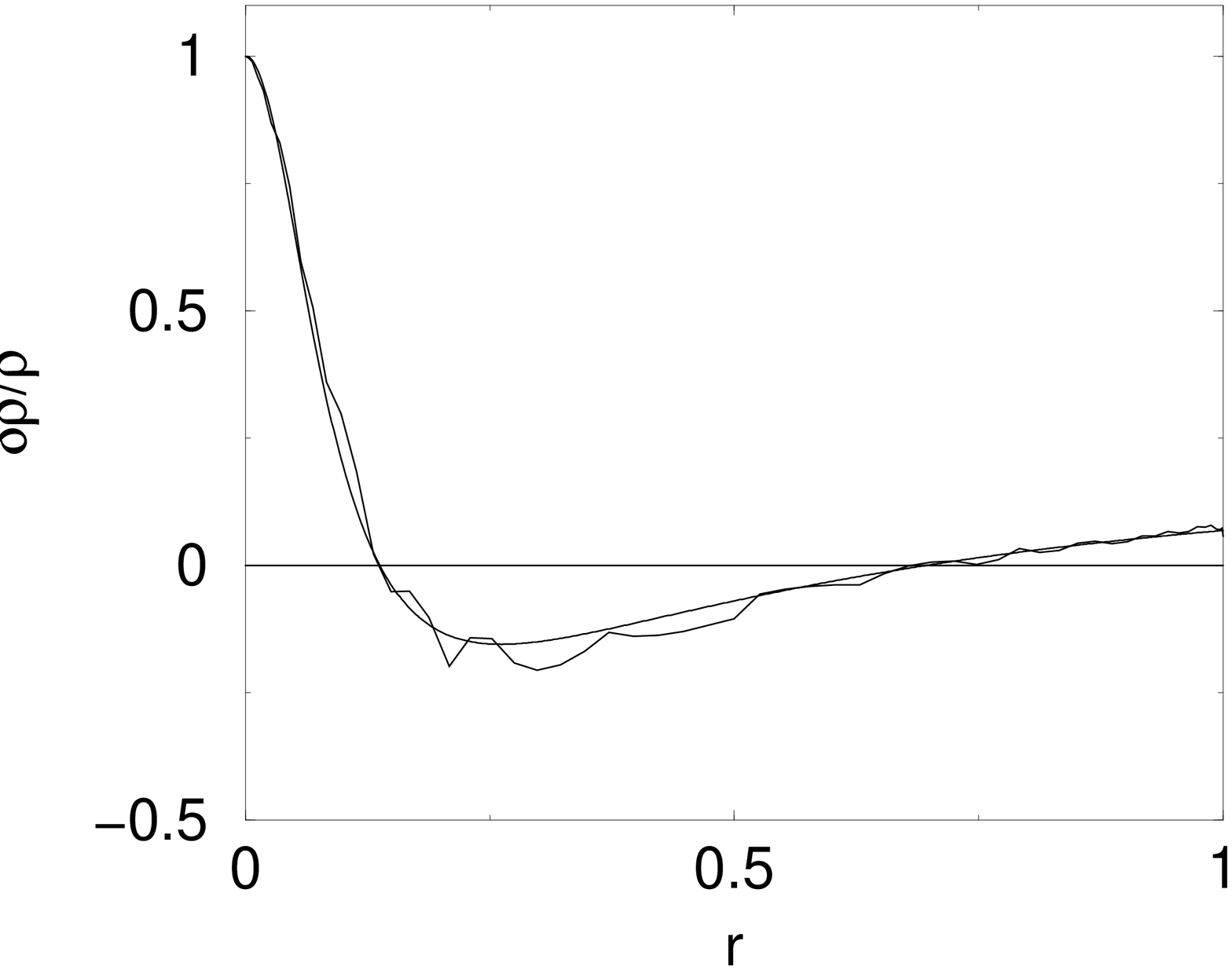,angle=-90,height=8.5cm}}
\caption{First mode of instability in the microcanonical ensemble.
The clean line is obtained by solving the eigenvalue equation
(\ref{l11}) with $\lambda=0$ and the broken line is obtained from the
numerical simulation of the SP system. The profile of density
perturbation presents a ``core-halo'' structure.} 
\label{dnsn-xL0.334417224}
\end{figure}

To study the development of the instability for short times, we start
from a point on the spiral of Fig.~\ref{pdiag} close to $\Lambda_{c}$
but with a density contrast ${\cal R}\gtrsim 709$ (we have taken
$\Lambda=0.3344$ and ${\cal R}=779$). This isothermal sphere, with
density profile $\rho_{eq}(r)$, is linearly unstable as it is a saddle
point of entropy (see Sec. \ref{sec_stab}). In
Fig.~\ref{dnsn-xL0.334417224}, we have represented the density
perturbation profile $\delta
\rho(r,t)/\rho_{eq}(r)=(\rho(r,t)-\rho_{eq}(r))/\rho_{eq}(r)$ that
develops for short times. This density profile presents a
``core-halo'' structure (i.e. it has two nodes) in excellent agreement
with the stability analysis of Padmanabhan \cite{pad2} (we have computed
the exact theoretical profile to compare quantitatively with the
simulation).

\subsection{Canonical ensemble}
\label{sec_cano}

\begin{figure}[htbp]
\centerline{
\psfig{figure=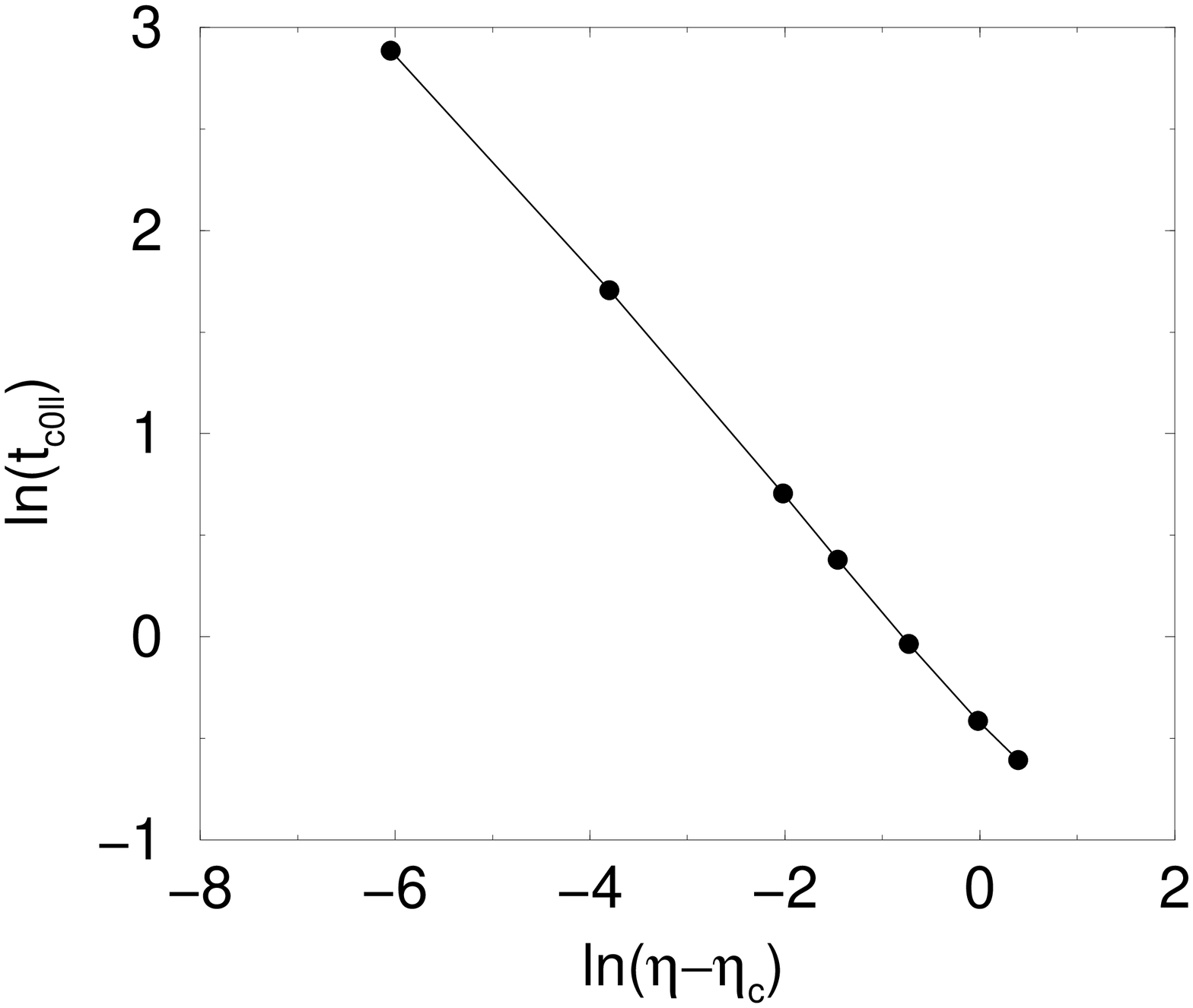,angle=-90,height=8.5cm}}
\caption{Evolution of the collapse time $t_{coll}$ with $\eta$. The
figure displays a scaling law
$t_{coll}\sim (\eta-\eta_{c})^{-\nu}$ with $\nu\sim 0.6$ close to the
theoretical value $1/2$.}
\label{tcollCANO}
\end{figure}

We now solve the SP system with a fixed temperature $T$. We
confirm the results of the thermodynamic approach in the canonical
ensemble. When $\eta<\eta_{c}$ the system converges to an
equilibrium state while it collapses for $\eta>\eta_{c}$
(isothermal collapse).  The collapse time $t_{coll}$ scales with
$\eta-\eta_{c}$ (see Fig.~\ref{tcollCANO}) with an exponent $\sim
-0.6$ close to the theoretical value $-1/2$.

\begin{figure}[htbp]
\centerline{
\psfig{figure=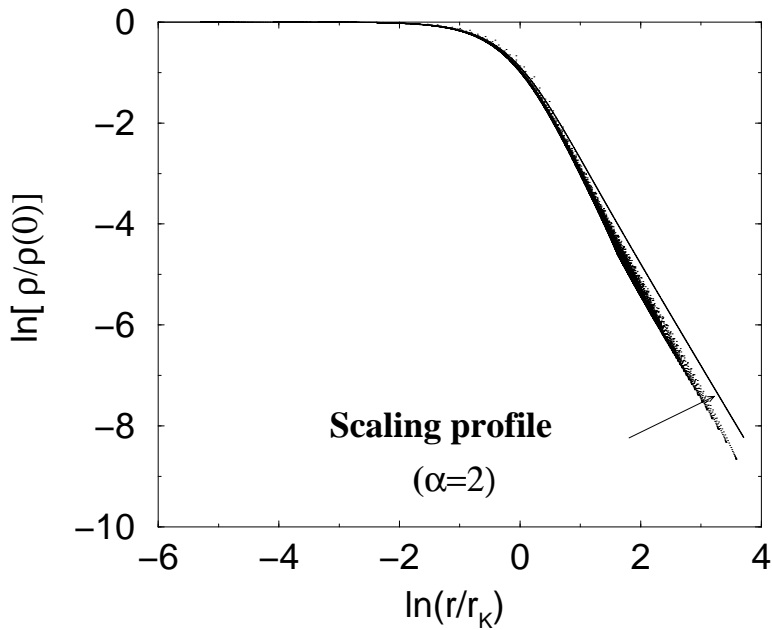,angle=-90,height=8.5cm}}
\caption{Self-similar profile for $\eta=2.75>\eta_{c}$. This
(quasi) invariant profile is compared with the analytical scaling
profile $f(x)$ with $\alpha=2$. Deviation from the pure scaling
law is due to non-scaling corrections that compensate for the
excess of mass contained in the scaling profile.}
\label{autosim2.75}
\end{figure}

\begin{figure}[htbp]
\centerline{
\psfig{figure=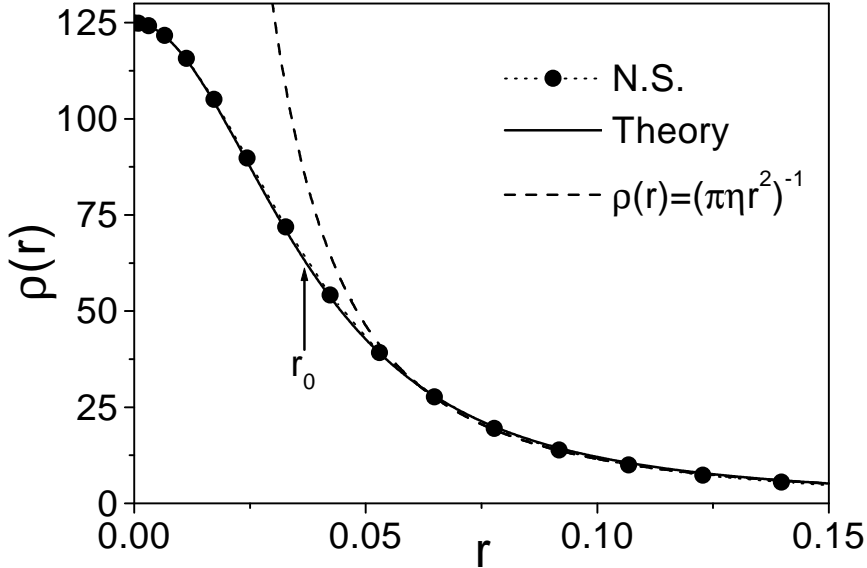,angle=0,height=8.5cm}}
\caption{We plot the numerical finite-time density profile for
$\eta=2.75$ (N.S.), at a time for which the central density is
$\rho(0,t)\approx 124.9\approx 28.8\rho_\infty$. This is compared to
the exact scaling profile $\rho_0f(r/r_0)$ (Theory), with $f$ given by
Eq.~(\ref{self10}), and $\rho_0=\frac{2\pi}{3}\rho(0,t)\approx 261.6$
and $r_0=(\eta\rho_0)^{-1/2}\approx 0.0373$
($\rho(r_0,t)/\rho(0,t)=14/27\approx 0.519$). We also plot the
asymptotic density profile, $\rho_{\rm as}=(\pi \eta r^2)^{-1}$, valid
for $r_0\ll r\ll 1$. In this region, the correction to scaling is
negligible.}
\label{collapse_fig1}
\end{figure}

\begin{figure}[htbp]
\centerline{
\psfig{figure=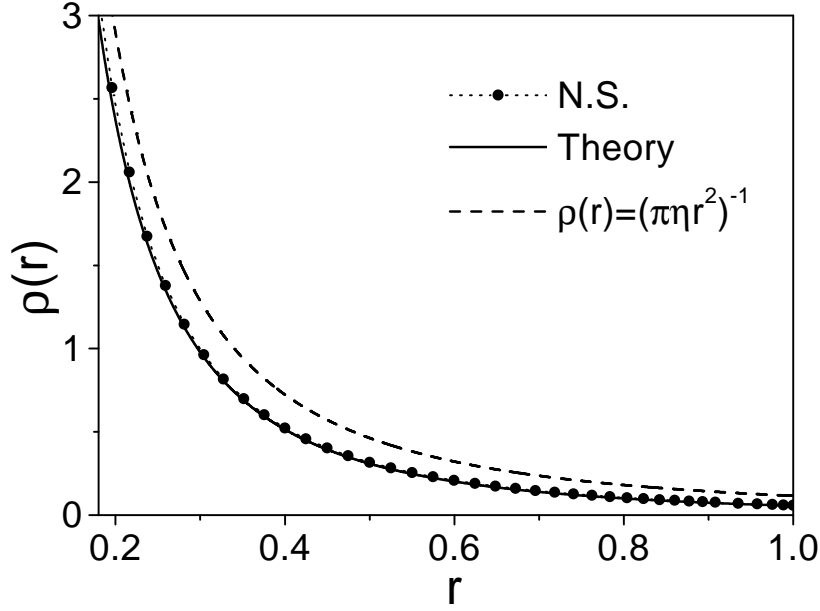,angle=0,height=8.5cm}}
\caption{We plot the same numerical data (N.S.) as in
Fig.~\ref{collapse_fig1}, but in the range $5r_0\leq r\leq 1$. This is
compared with the theoretical density profile at $t=t_{coll}$ obtained
from Eqs.~(\ref{an1}) and (\ref{self20}). The parameters $a\approx
5.0$ and $b\approx 5.1$ are determined by maximizing $\rho(1)$ (see
text), although the full profile barely depends on $a$ and $b$, as
soon as $b$ remains slightly greater than $a$, and $b\approx 4.8\sim
5.4$. In this range, the theoretical profile is in excellent agreement
with the numerical one. For instance, $\rho(1)_{\rm N.S.}\approx
0.058$ and $\rho(1)_{\rm Theory}\approx 0.057$. In order to stress the
quantitative agreement, we also plot the naive large $r$ asymptotics
of the scaling profile $\rho_{\rm as}=(\pi \eta r^2)^{-1}$, for which
$\rho(1)_{\rm as}\approx 0.116$.}
\label{collapse_fig2}
\end{figure}

\begin{figure}[htbp]
\centerline{
\psfig{figure=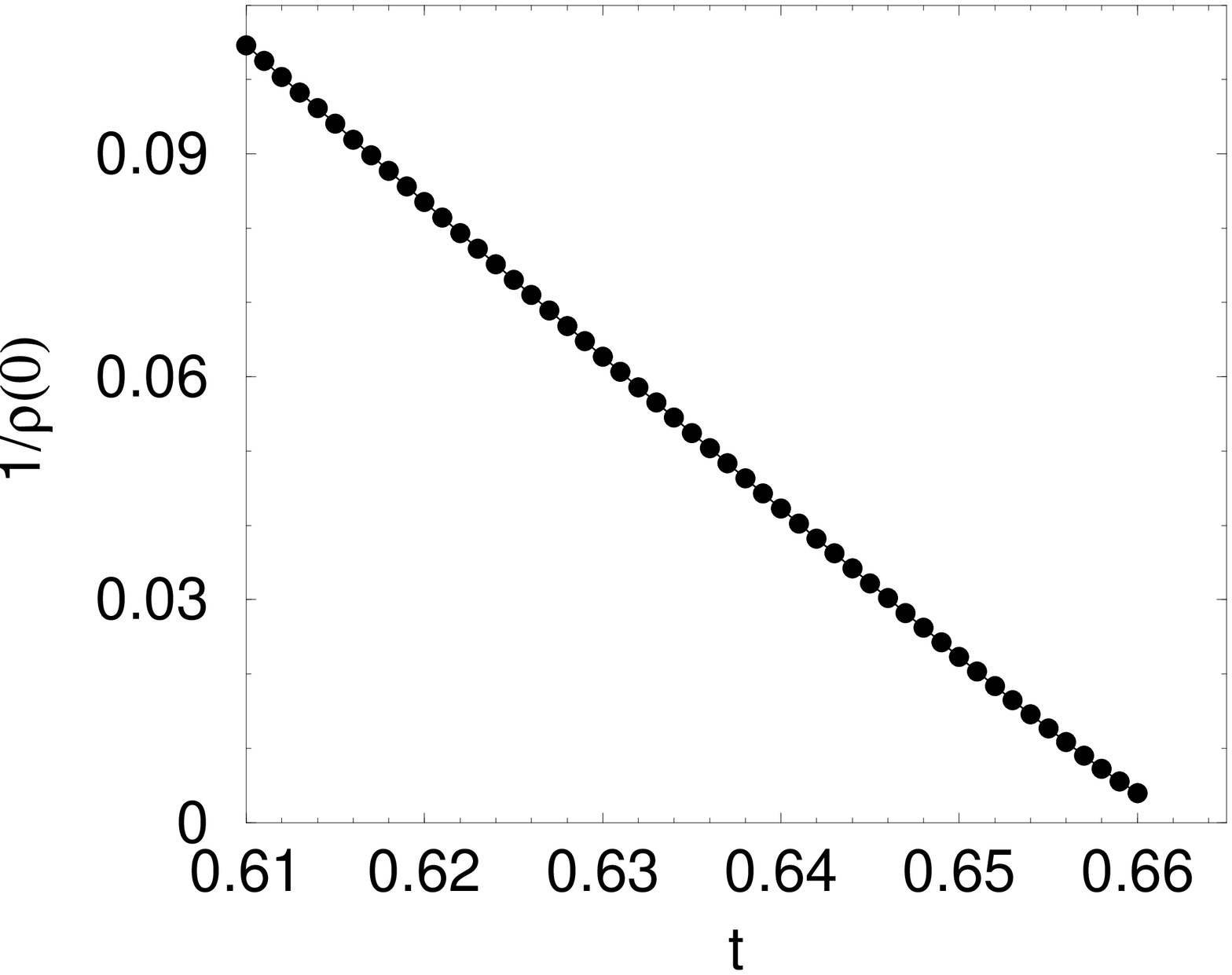,angle=-90,height=8.5cm}}
\caption{Time evolution of the inverse central density for
$\eta=3.5$. This curve displays a scaling regime
$1/\rho(0,t)=B(t_{coll}-t)$. The slope $B\simeq 2$ is close to the
theoretical prediction $2\pi/3\simeq 2.1$.} \label{beta3.5tcoll}
\end{figure}

In Fig.~\ref{autosim2.75}, we plot
the scaled density $\rho(r,t)/\rho(0,t)$ as a function of the scaled
distance $r/r_{K}(t)$ at different times. The curves tend to
superimpose but the thickness of the line indicates that we do not
have a strict self-similar regime (in agreement with our theoretical
analysis). Indeed, the invariant profile $f(x)$ computed in section
\ref{sec_sscano} matches the numerics very well in the core but does
not adequately describe the halo. The difference is due to the non
scaling part $F(r,t)$ that accounts for the mass conservation. In Figs.
\ref{collapse_fig1}-\ref{collapse_fig2}, the result of the numerical
simulation is compared more precisely with the full theoretical
prediction involving the non scaling term. The agreement is excellent
throughout the whole domain. In the core, the profile is dominated by
the scaling part which implies a $r^{-2}$ behavior at moderately large
distances. As explained previously and in Sec. \ref{sec_sscano},
this scaling behavior ceases to be valid near the wall and the
contribution of the non scaling part is clearly visible. Its influence
on the density profile remains weak but when the density is multiplied
by $r^{2}$, this non scaling profile has a non negligible contribution
to the total mass. In Fig.~\ref{beta3.5tcoll}, we see that the central
density diverges with time as $(t_{coll}-t)^{-1}$. The slope of the curve is
approximately equal to $2$ in good agreement with the theoretical
prediction $2\pi/3\simeq 2.1$ of section \ref{sec_sscano}.

\begin{figure}[htbp]
\centerline{
\psfig{figure=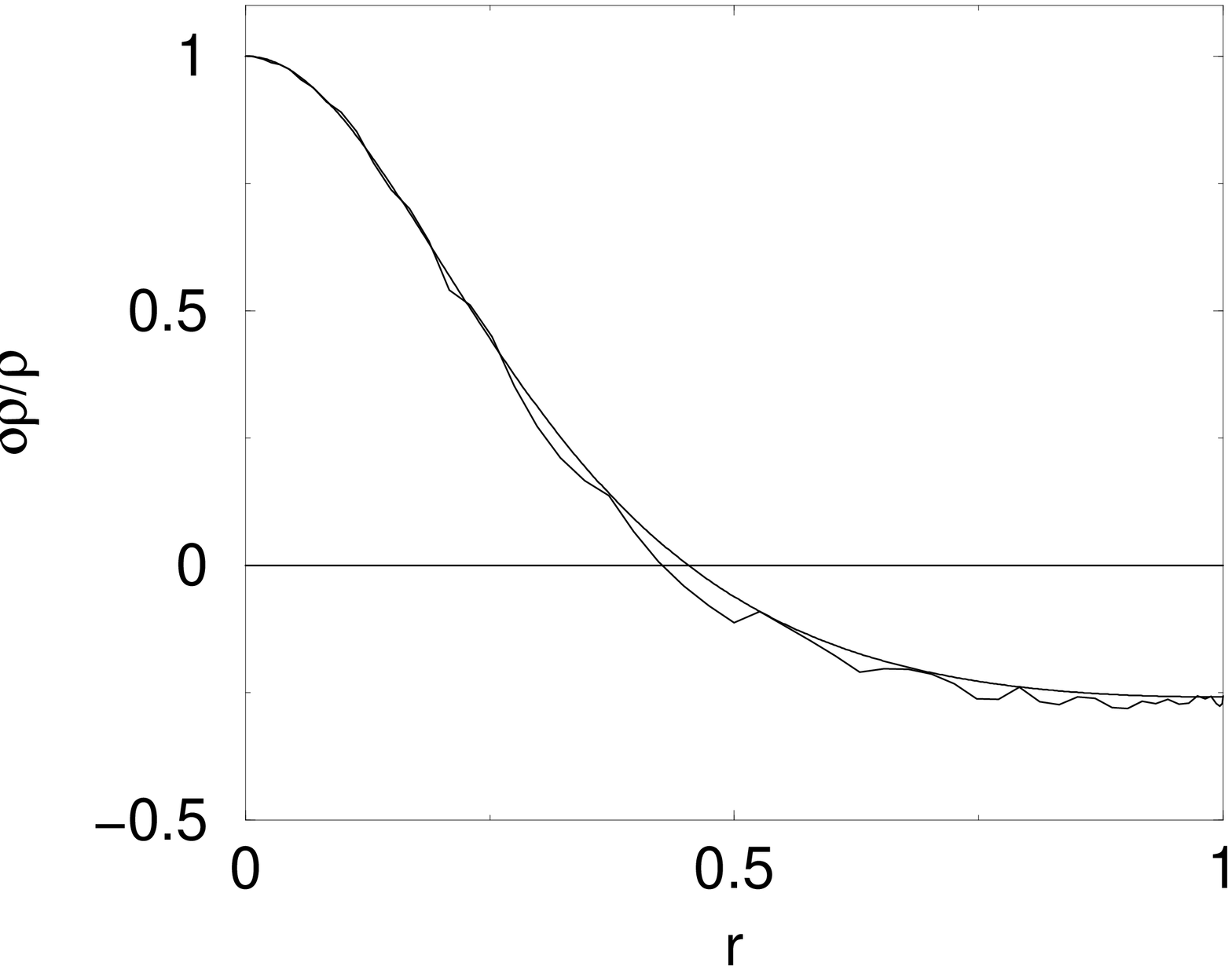,angle=-90,height=8.5cm}}
\caption{First mode of instability in the canonical ensemble. The
clean line is obtained by solving the eigenvalue equation (\ref{l13})
with $\lambda=0$ and the broken line is obtained from the numerical
simulation of the SP system. The density profile does not present a
``core-halo'' structure.} \label{dnsn-x.b2.51}
\end{figure}

\begin{figure}[htbp]
\centerline{
\psfig{figure=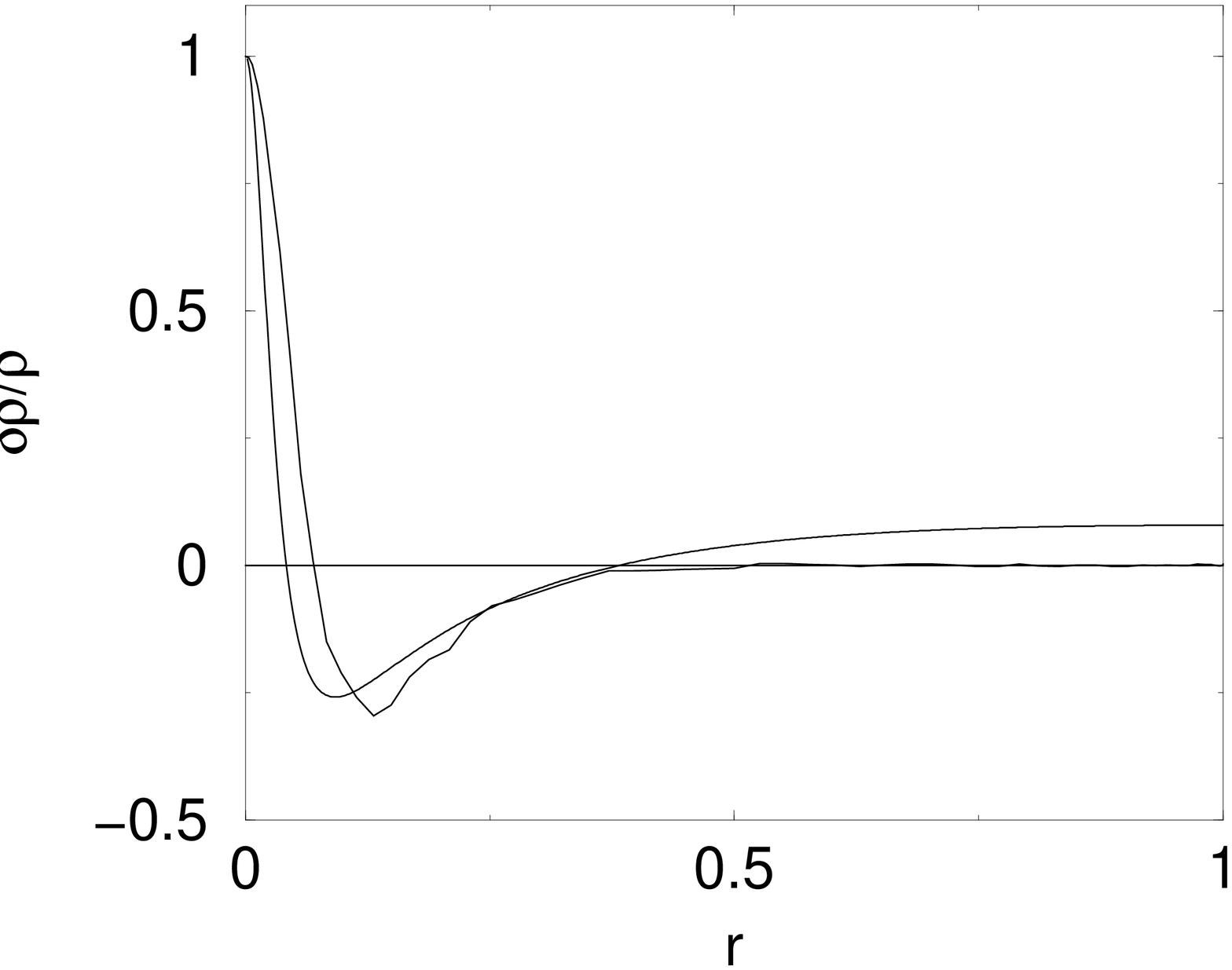,angle=-90,height=8.5cm}}
\caption{Second mode of instability in the canonical ensemble.}
\label{dnsn-xb1.84339333}
\end{figure}

In Fig.~\ref{dnsn-x.b2.51}, we study the early development of the
instability for $\eta\sim \eta_{c}$. More specifically, we start the
simulations from a point on the spiral of Fig.~\ref{pdiag} with
$\eta=2.510$ and ${\cal R}=42\gtrsim 32.1$. This isothermal sphere is
linearly unstable in the canonical ensemble as it is a saddle point of free
energy (see Sec. \ref{sec_stab}) and the perturbation profile that
develops for short times is shown in Fig.~\ref{dnsn-x.b2.51}. It is in
excellent agreement with the first mode of instability calculated by
Chavanis \cite{chavcano} in the canonical ensemble. This profile does
{\it not} present a ``core-halo'' structure, in
contrast with the first mode of instability in the microcanonical
situation. We have also plotted the perturbation profile for an
isothermal sphere located near the second extremum of temperature
($\eta=1.842...$) at which a new mode of instability appears
\cite{chavcano}. This second mode of instability has a core-halo
structure (Fig.~\ref{dnsn-xb1.84339333}). Of course, the perturbation
profile that develops is a superposition of the first two modes of
instability, but we see that its structure is dominated by the
contribution of the second mode.

In order to check the inequivalence of microcanonical and canonical
ensembles in the region of negative specific heats, we started the
simulation from an isothermal sphere with a density contrast comprised between
$32.1$ and $709$. In the first experiment, the energy is kept fixed
using the constraint (\ref{EE2}). In that case, it is found that the
sphere is linearly stable as it is a local entropy maximum. However,
if the temperature is fixed instead of the energy, the sphere is now
unstable as it is a saddle point of free energy. This clearly
demonstrates in the framework of our simple dynamical model that the
microcanonical and canonical ensembles are not interchangeable for
self-gravitating systems. This particular circumstance can be traced
back to the non-extensivity of the system due to the long-range nature
of the gravitational potential. This interesting problem is discussed
in the review of Padmanabhan \cite{pad} and illustrated by Chavanis
\cite{chavafermi} for specific models of self-gravitating systems
with a short-range cutoff (self-gravitating fermions and
hard-spheres models).

\begin{figure}[htbp]
\centerline{ \psfig{figure=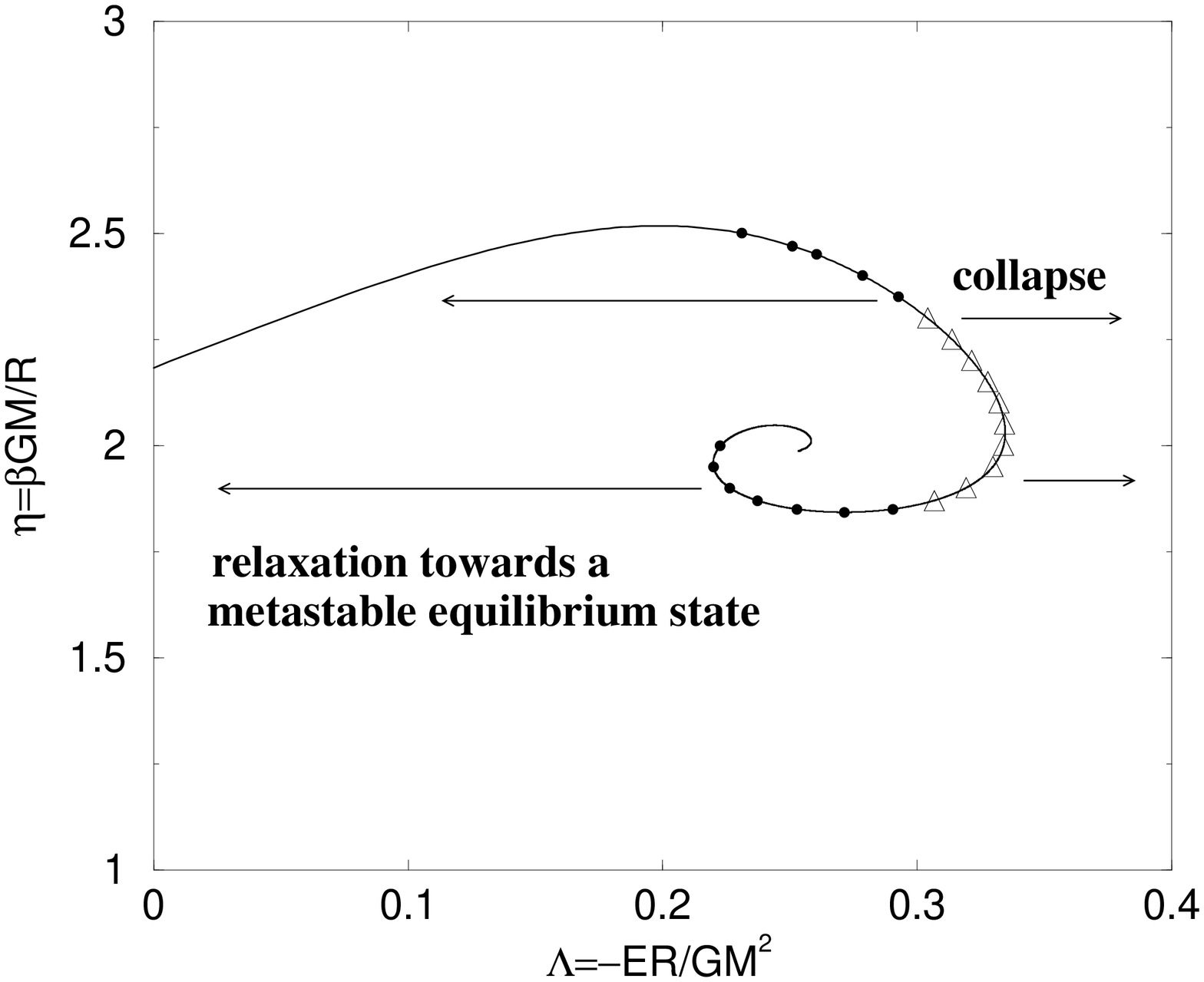,angle=-90,height=8.5cm}}
\caption{Basin of attraction in the canonical ensemble. The
isothermal spheres located after the first turning point of the
spiral are unstable in the canonical ensemble. Depending on their
position on the spiral (and the initial perturbation), they can
either relax towards the local maximum of free energy with same
temperature ($\bullet$) or undergo a gravitational collapse
($\triangle$).} \label{bassinT}
\end{figure}

Since the stable isothermal configurations are only {\it metastable}
(i.e., local maxima of a thermodynamical potential), the value of
energy or temperature is not sufficient to completely determine the
evolution of the system: depending on the {\it shape} of the density
profile, an initial configuration with $\Lambda<\Lambda_{c}$ or
$\eta<\eta_{c}$ can either reach a quiescent equilibrium state or
collapse. The actual evolution of the system depends whether the
initial configuration lies in the ``basin of attraction'' of the local
entropy maximum or not. Of course, the complete characterization of
this basin of attraction is an impossibly complicated task because we
would have to test all possible initial configurations. We have
limited our study in the canonical ensemble to the case of unstable
isothermal spheres located after the first turning point of
temperature. These solutions correspond to saddle points of free
energy. Therefore, a small perturbation (due here to numerical
roundoff error) can destabilize the system and induce a dynamical
evolution.  The question is whether the system evolves towards the
local maximum of free energy or undergoes gravitational
collapse. Since we start from a saddle point of free energy, the two
evolutions are possible depending on the form of the perturbation. In
addition, depending on the location of the saddle point on the spiral
(its density contrast), one of these evolutions may be preferred.  The
results of our study are displayed in Fig.~\ref{bassinT}. The
isothermal spheres that experienced a complete collapse in our
numerical experiments are marked with a symbol $\triangle$ while those
that converged towards an equilibrium state are marked with a symbol
$\bullet$. A kind of structure seems to emerge: it appears that the
isothermal spheres undergoing gravitational collapse in the canonical
ensemble are concentrated near the vertical tangent. We have found a
similar structure in the microcanonical ensemble with a concentration
of points undergoing gravitational collapse concentrated this time
near the lower horizontal tangent. However, as indicated previously,
this apparent structure is relevant at best in an average sense since
other initial perturbations of the {\it same} saddle point may lead to
a different evolution. In any case, these results confirm that the
maxima of entropy or free energy are not {\it global} maxima since
they do not attract all initial conditions.  While homogeneous spheres
with $\Lambda<\Lambda_{c}$ and $\eta<\eta_{c}$ always seem to converge
towards equilibrium, centrally concentrated systems with the same
control parameters can develop a self-similar collapse leading to a
finite time singularity. In fact, considering Fig.~\ref{bassinT}
again, we see that the central concentration is not the only condition
for collapse since there exists highly concentrated states that also
converge towards the smooth equilibrium profile with low density
contrast (in that case, the evolution corresponds to an
``explosion''). Therefore, the basin of attraction of the metastable
equilibrium states seems to have a highly non trivial structure. The
nonlinear stability of a linearly stable isothermal sphere (located
this time before the first turning point of energy or temperature) is
also of interest. Since it is not a global entropy maximum it can be
in principle destabilized by a finite amplitude perturbation. However,
this perturbation is expected to be huge so that, in practice, the
stability of the isothermal spheres with low density contrast is
extremely robust. This suggests that these metastable states can be
very long lived \cite{posch,miller,ispo} and physically relevant in an
astrophysical context.

\section{Conclusion}

This paper has discussed the thermodynamics and the collapse of a
system of self-gravitating Brownian particles in a high friction
limit. This approximation considerably simplifies the problem since
the evolution of the full distribution function $f({\bf r},{\bf v},t)$
is simply replaced by the evolution of its lowest moments. We showed
that the Smoluchowski-Poisson system presents a rich variety of
behaviors and displays interesting phase transitions between
equilibrium states and collapsing states depending on the value of
energy and temperature. When the two evolutions are possible, the
choice depends on a complicated notion of basin of attraction. This
simple model also illustrates dynamically the inequivalence of
statistical ensembles for systems with long-range interactions.

An extension of our study is to consider rotating systems with
conservation of angular momentum. The SP system can be generalized to
include rotation \cite{csr} and is interesting to study isothermal
configurations that are not spherically symmetric.  When spherical
symmetry is broken, it is possible that the system will fragment in
several clumps and that these clumps will themselves fragment in
substructures. This may yield a hierarchy of structures fitting one
into each other in a self-similar way as suggested by theoretical
considerations \cite{semelin,chavcano}. It would be of interest to
investigate whether the SP system can display a process of
fragmentation and exhibit a fractal behavior. Numerical simulations
are under way.

There exists a close analogy between the statistical mechanics of
self-gravitating systems and two-dimensional vortices
\cite{cthese,cfloride,japon}. Following the pioneering work of Onsager
\cite{onsager}, there has been some attempts to describe vortices as
maximum entropy structures, with possible applications to oceanic and
atmospheric situations (e.g., Jupiter's Great Red Spot). The
relaxation towards the maximum entropy state is usually described by a
Smoluchowski-Poisson system which analyzes the evolution of the
vorticity in terms of a diffusion and a drift. The diffusion is due to
the fluctuations of the velocity field and the drift to the
inhomogeneity of the vorticity field \cite{drift}. The SP system can
be deduced directly from the Liouville equation by using projection
operator technics \cite{ckinetic} or from a phenomenological maximum
entropy production principle \cite{rs}.  It is interesting to note
that, for point vortices, the Fokker-Planck equation directly has the
form of a Smoluchowski equation whereas for material particles this is
true only in a high friction limit. This is because, for point
vortices, the phase space coincides with the configuration space while
for material particles it involves the positions and the velocities of
the particles.

The Smoluchowski-Poisson system also appears in the description of
biological systems like bacterial populations \cite{murray}. The
diffusion is due to ordinary Brownian motion and the drift models a
chemically directed movement (chemotactic flux) along a concentration
gradient (of smell, infection, food,...). When the attractant
concentration is itself proportional to the bacterial density, this
results in a coupled system morphologically similar to the one studied
in the present paper. The question that naturally emerges is whether
this coupling can lead to an instability for bacterial populations
similar to the gravitational collapse of self-gravitating
systems. This possibility will be considered in a forthcoming paper in
which we consider self-similar solutions of the Smoluchowski-Poisson
equation for different systems in various space dimensions
\cite{scgravite}.

\section{Acknowledgments}
\label{sec_ack}

Preliminary results of this work were presented at the conference
on Multiscale Problems in Science and Technology (Dubrovnik, Sep
2000). One of us (P.H.C.) is grateful to W. Jaeger for mentioning
the connection of this work with biological systems.  This
research was supported in part by the National Science Foundation
under Grant No. PHY94-07194.

\newpage
\appendix

\section{Analytical study of the scaling equation}
\label{sec_scalingeq}

In this Appendix, we study analytically the scaling equation
(\ref{self8}). To that purpose, we rewrite it in an equivalent albeit
more convenient form. Let us introduce the function
\begin{equation}
g(x)=4\pi\int_{0}^{x}f(x')x^{'2}dx',
\label{Q1}
\end{equation}
in terms of which Eq. (\ref{self8}) becomes
\begin{equation}
f(x)+{x\over\alpha}f'(x)={1\over x^{2}}{d\over dx} \lbrace x^{2}f'(x)+f(x)g(x)\rbrace.
\label{Q2}
\end{equation}
Multiplying both sides of equation (\ref{Q2}) by $x^{2}$ and
integrating the resulting expression between $0$ and $x$, we obtain
\begin{equation}
g(x)= 4\pi x^{2}{xf(x)-\alpha f'(x)\over 3-\alpha +4\pi \alpha f(x)}.
\label{Q3}
\end{equation}
From Eqs. (\ref{Q1}) and (\ref{Q3}), we can derive a nonlinear
recursion relation satisfied by the coefficients $a_{n}$ of the series
expansion of $f(x)$ in powers of $x^{2}$ (as $f$ is an even
function). Writing
\begin{equation}
f(x)={1\over 4\pi}\sum_{n=0}^{+\infty}(-1)^{n}a_{n}x^{2n},
\label{Q5}
\end{equation}
we find
\begin{equation}
a_{n+1}=-{2n+\alpha\over 2\alpha(n+1)(2n+3)}a_{n}+{1\over
2(n+1)}\sum_{p=0}^{n}{a_{p}a_{n-p}\over 2p+3}.
\label{Q6}
\end{equation}
This recursion relation leads to the large $n$ behavior of $a_{n}$:
\begin{equation}
a_{n}\sim 8r\biggl (n+{3\over 2}\biggr )r^{n}+o(r^{n}),
\label{Q7}
\end{equation}
where $r$ is an unknown constant related to the inverse radius of
convergence of the series. For $\alpha=2$, the asymptotics given by
Eq. (\ref{Q7}) with $r=1/2$ is an {\it exact} solution of the
recursion relation (\ref{Q6}), as can be checked by direct
substitution. Using the identities
\begin{equation}
P(x)={1\over 1+r x^{2}}=\sum_{n=0}^{+\infty} (-1)^{n}r^{n}x^{2n},
\qquad P'(x)=-{2rx\over (1+r x^{2})^{2}}={2\over
x}\sum_{n=0}^{+\infty}(-1)^{n}nr^{n}x^{2n}, \label{Q8}
\end{equation}
the series (\ref{Q5}) can easily be resummed
leading to Eq. (\ref{self10}).

\section{The case of cold systems ($T=0$)}
\label{sec_cold}

For $T=0$, the core radius is not given by the King radius (\ref{self2})
which is zero by definition. We still assume however that
$\rho_{0}\overline{r}_{0}^{\alpha}\sim 1$, where $\alpha$ is unknown
{\it a priori}. The equation for the invariant profile is then given
by
\begin{equation}
f(x)+{x\over\alpha}f'(x)={1\over x^{2}}{d\over dx}(f(x)g(x)),
\label{S24}
\end{equation}
where $g(x)$ is defined by Eq. (\ref{Q1}).  Multiplying
Eq. (\ref{S24}) by $4\pi x^{2}$ and integrating from $0$ to $x$ we obtain
\begin{equation}
g(x)={4\pi x^{3}f(x)\over 3-\alpha+4\pi \alpha f(x)}.
\label{S26}
\end{equation}
Using the relation $f(x)=g'(x)/4\pi x^{2}$, the foregoing equation can
be rewritten
\begin{equation}
(\alpha-3)g(x)+x g'(x)=\alpha {1\over x^{2}} g'(x)g(x).
\label{S26a}
\end{equation}
Introducing the change of variables $u=x^{3}$, we get
\begin{equation}
3{dg\over du}={(3-\alpha)g\over u-\alpha g}.
\label{S26b}
\end{equation}
A separation of the variables can be effected by the transformation
$g=u h$, yielding
\begin{equation}
{1-\alpha h\over h(3h-1)}dh={\alpha\over 3}{du\over u}.
\label{S26c}
\end{equation}
This equation is readily integrated leading to the implicit equation
\begin{equation}
g(x)=\lambda \biggl ({x^{3}\over 3}-g(x)\biggr )^{1-\alpha/3},
\label{S26d}
\end{equation}
where $\lambda$ is an integration constant.  As $g(x)$ is an odd
analytical function, Eq.~(\ref{S26d}) first implies that $g(x)\sim
{x^{3}\over 3}$, so that $f(0)={1\over 4\pi}$. Combining with Eq.
(\ref{self7}), this yields $\rho(0,t)={\xi\over 4\pi
G}(t_{coll}-t)^{-1}$. Then, inserting $g(x)-{x^{3}\over 3}\sim
x^{5}$ in Eq.~(\ref{S26d}), we find that $x^{3}\sim
x^{5(1-\alpha/3)}$, leading to $\alpha=6/5$. Note finally that the
scaling profile defined by the implicit equation (\ref{S26d})
 can be written in the parametric form
\begin{equation}
f(x)={1\over 4\pi}{1\over 1+s}, \qquad g(x)={1\over
3}s^{3/2},\qquad x=s^{1/2}\biggl (1+{3\over 5}s\biggr )^{1/3},
\label{S30}
\end{equation}
where the constant $\lambda$ has been incorporated in the
expression of the  core radius $\overline{r}_{0}$.

In fact, for $T=0$,  Eq. (\ref{X5}) can be solved analytically.
Since the diffusion term vanishes, this equation describes a {\it
deterministic} motion where the particles have a velocity ${\bf
u}=-{1\over\xi}\nabla\Phi$ directly proportional to the
gravitational force (see Sec. \ref{sec_smoluchowski}). This
deterministic problem can be solved exactly by adapting the
procedure followed by Penston \cite{penston} in his investigation
of the collapse of cold self-gravitating gaseous spheres. Let us
consider a particle located at $r(0)=a$ at time $t=0$. We denote
by $\overline{\rho}(a)$ the average density inside the sphere of
radius $a$. The total mass inside radius $a$ can therefore be
expressed as $M_{a}={4\pi\over 3}\overline{\rho}(a)a^{3}$. At time
$t$, this mass is now contained in the sphere of radius $r=r(t)$,
where $r(t)$ is the position of the particle initially at $r=a$.
Using the Gauss theorem, the motion of the particle is described
by the first order differential equation
\begin{equation}
{dr\over dt}=-{1\over\xi}{GM_{a}\over r^{2}}.
\label{S3}
\end{equation}
This equation can be integrated explicitly to give
\begin{equation}
r=a\biggl (1-{4\pi G\over\xi}\overline{\rho}(a)t\biggr )^{1/3}.
\label{S4}
\end{equation}
Let us first discuss the case where the system is initially
homogeneous with density
$\overline{\rho}(a)=\overline{\rho}_{0}$. In that case, all the particles
(whatever their initial position) arrive at $r=0$ at a time
$t_{coll}= {\xi/ 4\pi G\overline{\rho}_{0}}$
defined as the collapse time for $T=0$. This expression represents a
lower bound (reached for $\eta\rightarrow +\infty$) on the value of the
collapse time $t_{coll}(\eta)$ studied in Sec.
\ref{sec_critphen}. During  the evolution, the sphere
remains homogeneous with radius, density and free energy evolving as
\begin{equation}
R(t)=R(1-t/t_{coll})^{1/3},\quad
\rho(t)=\overline{\rho}_{0}(1-t/t_{coll})^{-1},\quad J(t)={3\beta
GM^{2}\over 5R}(1-t/t_{coll})^{-1/3}. \label{S6}
\end{equation}
Note that the free energy diverges at $t=t_{coll}$, unlike in
Sec. \ref{sec_sscano}. These results can also be obtained directly
from Eq.  (\ref{X5}) which reduces, for a uniform density, to
\begin{equation}
{d\rho\over dt}=\nabla\biggl ({1\over\xi}\rho\nabla\Phi\biggr
)={1\over\xi}\rho\Delta\Phi={4\pi G\over\xi} \rho^{2},
\label{S7}
\end{equation}
where we have used the Poisson equation (\ref{Poisson}) to get the
last equality.  

We now suppose that, initially, $\overline{\rho}(a)$ has a smooth
maximum at the center so that
\begin{equation}
\overline{\rho}(a)=\overline{\rho}_{0}\biggl (1-{a^{2}\over A^{2}}\biggr ),
\label{S8}
\end{equation}
for sufficiently small $a$, where $A$ is a constant. In that case,
Eq. (\ref{S4}) giving the position at time $t$ of the particle
located at $r=a$ at $t=0$ becomes
\begin{equation}
r=a\biggl \lbrack 1-\biggl (1-{a^{2}\over A^{2}}\biggr ){t\over
t_{coll}}\biggr \rbrack^{1/3}.
\label{S9}
\end{equation}
At $t=t_{coll}$, the time at which the central density becomes
infinite, it reduces to $r={a^{5/3}/ A^{2/3}}$. It is now
straightforward to obtain the full density profile at
$t=t_{coll}$. Since the mass contained between $a$ and $a+da$ at
$t=0$ arrives between $r$ and $r+dr$ at time $t$, we have in full
generality
\begin{equation}
\overline{\rho}(a)4\pi a^{2}da =\rho(r,t)4\pi r^{2}dr,
\label{S11}
\end{equation}
 or, for sufficiently small $a$,
\begin{equation}
\rho(r,t)=\overline{\rho}(a){a^{2}\over r^{2}}{da\over dr}\simeq
\overline{\rho}_{0}{a^{2}\over r^{2}}{da\over dr}.
\label{S12}
\end{equation}
At $t=t_{coll}$, we get
\begin{equation}
\rho(r,t_{coll})={3\over 5}\overline{\rho}_{0}A^{6/5}r^{-6/5}.
\label{S13}
\end{equation}
We have therefore recovered that, for $T=0$, the density profile
decreases algebraically with an exponent $\alpha=6/5$. We now
extend this analysis to a time $\tau=t_{coll}-t$ just before the
singularity arises. Considering the limit $a\rightarrow 0$ and
$\tau\rightarrow 0$, Eq. (\ref{S9}) can be expanded to lowest
order as
\begin{equation}
r=a\biggl ({\tau\over t_{coll}}+{a^{2}\over A^{2}}\biggr )^{1/3}.
\label{S14}
\end{equation}
Then, Eq. (\ref{S12}) leads, after some reductions, to the density profile
\begin{equation}
\rho(r,t)={\overline{\rho}_{0}\over {\tau\over t_{coll}}+{5a^{2}\over 3A^{2}}}.
\label{S15}
\end{equation}
The central density corresponds to $r=0$, i.e. $a=0$. According to
Eq. (\ref{S15}) it evolves with time as
\begin{equation}
\rho(0,t)={\overline{\rho}_{0}t_{coll}\over\tau}=
{\xi\over 4\pi G}(t_{coll}-t)^{-1}.
\label{S16}
\end{equation}
Therefore, if we define
\begin{equation}
s={5a^{2}t_{coll}\over 3A^{2}\tau},\qquad \overline{r}_{0}=\biggl
({3A^{2}\over 5}\biggr )^{1/2} \biggl ({\tau\over t_{coll}}\biggr
)^{5/6}. \label{S17}
\end{equation}
we can express the density profile in the parametric form
\begin{equation}
{\rho(r,t)\over\rho (0,t)}={1\over 1+s},\qquad  {r\over
\overline{r}_{0}(t)}=s^{1/2}\biggl (1+{3\over 5}s\biggr )^{1/3},
\label{S18}
\end{equation}
which is equivalent to Eq. (\ref{S30}).
According to Eq. (\ref{S16}) and (\ref{S17}), we have the scaling laws
$\overline{r}_{0}\sim (t_{coll}-t)^{5/6}$,
$\rho(0)\overline{r}_{0}^{6/5}\sim 1$
just before the singularity occurs. Setting $F=\rho/\rho(0)$ and
$x=r/\overline{r}_{0}$, we  easily check that $F(x)=1-x^{2}+...$ for
$x\rightarrow 0$ and $F(x)\sim ({3/5})^{2/5} x^{-6/5}$ for
$x\rightarrow +\infty$.
This solves the problem for $T=0$. Now, if the temperature $T$ is very
small but non-zero, we expect the present scaling to hold provided
that $\overline{r}_{0}\gg r_{0}(t)$, where $r_{0}$ is defined in
section \ref{sec_selfsimilar}. This leads to a cross-over core density
$\rho_{0}^{*}$ above which the $T\neq 0$ scaling of section
\ref{sec_sscano} will prevail. The density $\rho_{0}^{*}$ can be
estimated by equating $r_{0}=(T/G\rho_{0})^{1/2}$ to
$\overline{r}_{0}\sim \rho_{0}^{-5/6}$. The $T\neq 0$ scaling then
prevails when the density becomes high enough, $\rho_{0}^{*}\sim ({T/G})^{-3/2}$.

\section{Connection between dynamical and thermodynamical stability}
\label{sec_connexion2}

Let $\rho$ be a stationary solution of Eq. (\ref{X5}) and
$\delta \rho$ a small perturbation around this solution. The first and
second variations of temperature respecting the energy constraint
(\ref{EE2}) can be expressed as
\begin{equation}
{3\over 2}M\delta T+\int \delta \rho \Phi \,d^{3}{\bf r}=0,
\label{deltaT}
\end{equation}
\begin{equation}
{3\over 2}M\delta^{2} T+{1\over 2}\int \delta \rho \delta \Phi \,d^{3}{\bf r}=0.
\label{delta2T}
\end{equation}
The critical point $\rho$ is a local entropy {\it maximum} provided that the
second variations of entropy
\begin{equation}
\delta^{2}S=-{3 M\over 4}{(\delta T)^{2}\over T^{2}}+{3M\over 2}
{\delta^{2}T\over T}-{1\over 2}\int {(\delta \rho)^{2}\over \rho}\,d^{3}{\bf r}
\label{delta2Jbis}
\end{equation}
are negative for any variations that conserve mass to first order. Let
us now linearize Eq. (\ref{X5}) around equilibrium and write
the time dependence of the perturbation in the form $\delta \rho\sim
e^{\lambda t}$. We get
\begin{equation}
\lambda\delta \rho=\nabla \biggl \lbrack {1\over\xi}(\delta T\nabla
\rho +T \nabla \delta \rho+\delta \rho\nabla \Phi+
\rho\nabla\delta\Phi)\biggr\rbrack.
\label{man1}
\end{equation}
Multiplying both sides of Eq. (\ref{man1}) by $\delta \rho/\rho$,
integrating by parts and using the equilibrium condition
$T\nabla\rho+\rho\nabla\Phi=0$, we obtain
\begin{equation}
\lambda\int {(\delta \rho)^{2}\over \rho}\,d^{3}{\bf r}=
-\int {1\over T \rho\xi} (T\nabla \delta \rho+\delta \rho\nabla \Phi)
(\delta T\nabla \rho+T\nabla \delta \rho+\delta \rho\nabla
\Phi+\rho\nabla\delta\Phi)\,d^{3}{\bf r}.
\label{man2}
\end{equation}
We now remark that the second order variations of the rate of entropy
production (\ref{dotSeq}) are given by
\begin{equation}
\delta^{2}\dot S=\int {1\over \rho T\xi}(\delta T\nabla \rho+T\nabla
\delta \rho+\delta \rho\nabla \Phi+\rho\nabla\delta\Phi)^{2}\,d^{3}{\bf r}.
\label{delta2Sbis}
\end{equation}
We can therefore rewrite Eq. (\ref{man2}) in the form
\begin{eqnarray}
\lambda\int {(\delta \rho)^{2}\over \rho}\,d^{3}{\bf r}=-\delta^{2}\dot S
+\int {1\over T \rho\xi} (\delta T\nabla \rho+\rho\nabla\delta\Phi)\nonumber\\
\times (\delta T\nabla\rho +T\nabla \delta \rho+\delta \rho\nabla \Phi+
\rho\nabla\delta\Phi)\,d^{3}{\bf r}.
\label{man3}
\end{eqnarray}
Using the equilibrium condition, the last term in Eq. (\ref{man3})
is clearly the same as
\begin{equation}
-\int {1\over\xi} (\delta T\nabla \rho+T\nabla \delta \rho+\delta
\rho\nabla \Phi+\rho\nabla\delta\Phi) \biggl ({\delta T\over
T^{2}}\nabla\Phi-{1\over T}\nabla\delta\Phi\biggr ) \,d^{3}{\bf r}.
\label{man4}
\end{equation}
Taking the time derivative of Eq. (\ref{EE2}) and using
Eq. (\ref{X5}) we have at each time
\begin{equation}
\dot E={3\over 2}M\dot T-\int{1\over\xi}(T\nabla \rho+\rho\nabla\Phi)
\nabla\Phi \,d^{3}{\bf r}=0.
\label{dotE2}
\end{equation}
The energy constraint (\ref{dotE2}) must be satisfied to first and
second order. This yields:
\begin{equation}
\int {1\over\xi} (\delta T\nabla \rho+T\nabla \delta \rho+
\delta \rho\nabla \Phi+\rho\nabla\delta\Phi)\nabla\Phi
\,d^{3}{\bf r}={3\over 2}M\delta \dot{T}={3\over 2}M\lambda \delta {T},
\label{cons1}
\end{equation}
\begin{equation}
\int {1\over\xi} (\delta T\nabla \rho+T\nabla \delta \rho+\delta
\rho\nabla \Phi+\rho\nabla\delta\Phi)\nabla\delta \Phi
\,d^{3}{\bf r}={3\over 2}M\delta^{2} \dot{T}={3}M\lambda \delta^{2} {T},
\label{cons2}
\end{equation}
where we have used Eqs. (\ref{deltaT})-(\ref{delta2T}) to obtain
the last equalities. Substituting these relations in Eq.
(\ref{man3}), we get
\begin{equation}
\lambda\biggl \lbrace \int {(\delta \rho)^{2}\over \rho}\,d^{3}{\bf r}+
{3M\over 2}{(\delta T)^{2}\over T^{2}}-3M{\delta^{2}T\over T}
\biggr\rbrace=-\delta^{2}\dot S.
\label{man5}
\end{equation}
Comparing with Eq. (\ref{delta2Jbis}), we finally obtain
\begin{equation}
\delta^{2}\dot S=2\lambda\delta^{2}S.
\label{man6}
\end{equation}
Since $\delta^{2}\dot S\ge 0$, see Eq. (\ref{delta2Sbis}), the sign of
$\lambda$ is the same as that of $\delta^{2}S$. If $\rho$ is a local
entropy maximum, then $\delta^{2}S$ and consequently $\lambda$ are
negative for any perturbation: the solution is linearly
stable. Otherwise, we can find a perturbation for which $\delta^{2}S$,
and consequently $\lambda$, are positive: the solution is linearly
unstable. We can easily extend the relation (\ref{man6}) to the
canonical ensemble with $J$ instead of $S$. We have found the same
relation for other types of kinetic equations (Chavanis, in preparation), so
its validity seems to be of a very wide scope.

\end{document}